**Virtual Reality for Emotion Elicitation – A Review**

Rukshani Somarathna, School of Computer Science and Engineering, University of New South Wales, Australia r.somarathna@student.unsw.edu.au

Tomasz Bednarz, School of Art and Design, School of Computer Science and Engineering, University of New South Wales, Australia t.bednarz@unsw.edu.au

Gelareh Mohammadi, School of Computer Science and Engineering, University of New South Wales, Australia g.mohammadi@unsw.edu.au

Corresponding author - Rukshani Somarathna (r.somarathna@student.unsw.edu.au)

**Abstract**

Emotions are multifaceted phenomena that affect our behaviour, perception, and cognition. Increasing evidence indicates that induction mechanisms play a crucial role in triggering emotions by simulating the sensations required for an experimental design. Over the years, many reviews have evaluated a passive elicitation mechanism where the user is an observer, ignoring the importance of self-relevance in emotional experience. So, in response to the gap in the literature, this study intends to explore the possibility of using Virtual Reality (VR) as an active mechanism for emotion induction. Furthermore, for the success and quality of research settings, VR must select the appropriate material to effectively evoke emotions. Therefore, in the present review, we evaluated to what extent VR visual and audio-visual stimuli, games, and tasks, and 360-degree panoramas and videos can elicit emotions based on the current literature. Further, we present public datasets generated by VR and emotion-sensing interfaces that can be used in VR based research. The conclusions of this survey reveal that VR has a great potential to evoke emotions effectively and naturally by generating motivational and empathy mechanisms which makes it an ecologically valid paradigm to study emotions.

**Keywords**: Emotions, Emotion induction, Virtual Reality, Elicitation, Affective Computing

## 1 Introduction

Emotions are a significant part of human associations correlated with nonverbal communication [1], behaviour [2], physical and mental changes [3]. It is a sensory condition associated with cognition, behaviour, and arousal that leads to physical and psychological changes [4, 5]. The conceptualization of emotional definition is complicated by the broader mechanism involved in the phenomenon and although, there is no consensus on the definition of emotion, the following factors can be interpreted without debate: involvement of several components of the human body, response based on subjective evaluation, and readiness to deal with stimuli.

Neuroscience portrayed the role of emotions in human creativity [6], cognition [7, 8], decision making [9-11], and brain activity [12, 13], and therefore it is important for computers to have the expertise to identify and express emotions to facilitate Human-Computer Interactions (HCI) [6]. At the same time, with the focus on Artificial Intelligence (AI), Affective Computing (AC) emerged as a computing paradigm that concerns emotion modeling, recognition, and synthesis. As conceptualized by Paiva, et al. [14], the Affective Loop of emotive machines and users consists of emotional elicitation, recognition, and behaviour generation (expression, adaptation, and synthesis) as seen in Figure 1. This emphasizes the gravity of understanding emotions in HCI and underscores the significant need to understand the underlying mechanisms of emotions. It is essential in the affective science to design an effective emotional environment to create ecologically valid affective statuses to obtain reliable results. This often depends on successful emotion induction [15].

Multidisciplinary science has used numerous emotion induction mechanisms to evoke a range of emotions in controlled experiments which can be broadly grouped into passive and active. However, passive methods are a more pronounced source compared to active methods which are more realistic [16]. Although many studies have successfully manipulated emotions using active mechanisms, to date, there is still controversy over the most optimal mechanism for creating real feelings in laboratory settings. The use of Virtual Reality (VR) has witnessed remarkable growth in psychological studies in recent years. Therefore, we hypothesize VR as an effective and powerful emotional induction mechanism to generate a variety of emotions and empathy machines [17]. Although VR has already proven to be an appropriate medium for emotion elicitation, it is important to also highlight its limitations and ways to improve its quality.

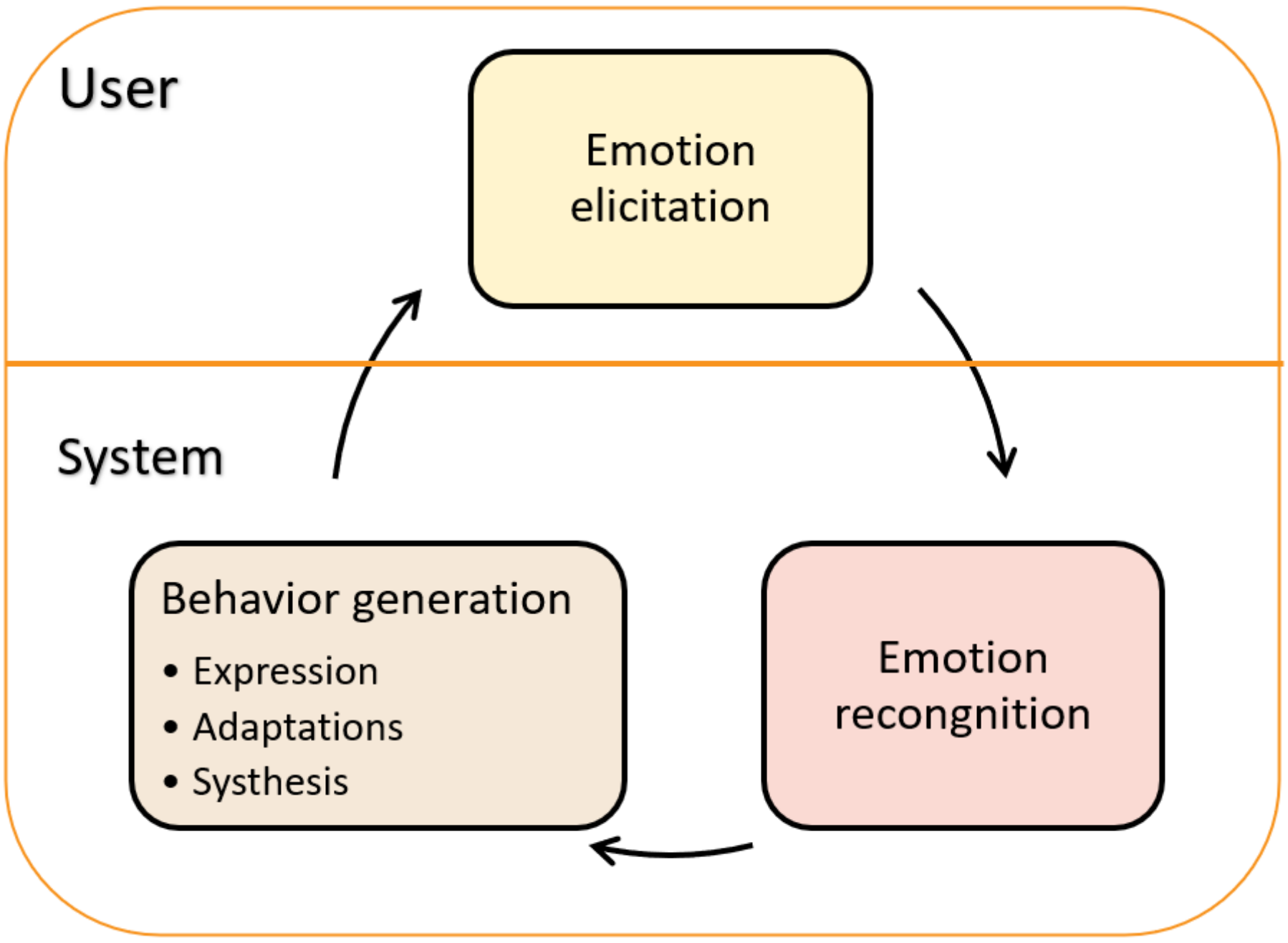

Figure 1 Affective Loop theory. Figure based on [14] .

Several studies have used VR as a major stimulant [18-22], but survey publications rarely mention [23] or did not refer [24] to it as an elicitation material. Also, the exact mechanism of manipulating interactive content to generate specific emotional experiences has been debated [25, 26]. Inspired by the lack of a comprehensive empirical study, the purpose of our review is to assess the potential of VR in the emotion-inducing process which is required for laboratory experimental conduct. In consonance with that, this study investigates the past paradigm of literature and makes recommendations for selecting the materials for emotion elicitation. Data sources are a mandatory requirement for psychological research and the establishment of a database can be quite cumbersome and expensive. Therefore, in our survey, we will highlight publicly available datasets generated using VR media content and further about applicable subjective and objective measures in establishing a novel dataset. Moreover, the findings of the analysis will be directed to the formulation of a more rigorous and successful research framework. These will lead to more realistic elicitation of emotion, which are important for detection, synthesis, and analysis of the underlying mechanisms of emotions.

In consonance with the statement that different methods generate different intensities of emotions [24, 27], and after an analysis of publications that effectively induced emotions in the literature, we classify VR based media content into four main categories including visual or audio-visual stimuli, games and tasks, 360-degree media, and mixed reality as shown in Figure 2. Moreover, we explored the number of research papers that incorporated VR in emotional studies. Figure 3 shows the upward trend of VR usage over the past decade from 2011 to 2021. The results were obtained from Google Scholar, searching the keywords "Virtual Reality" AND ("Affective computing" OR Emotions) AND (Elicitation OR Induction) anywhere in the article. The graph clearly shows that literature has an increasing interest in the utilization of VR to study emotions or to utilize emotions in interaction designs.

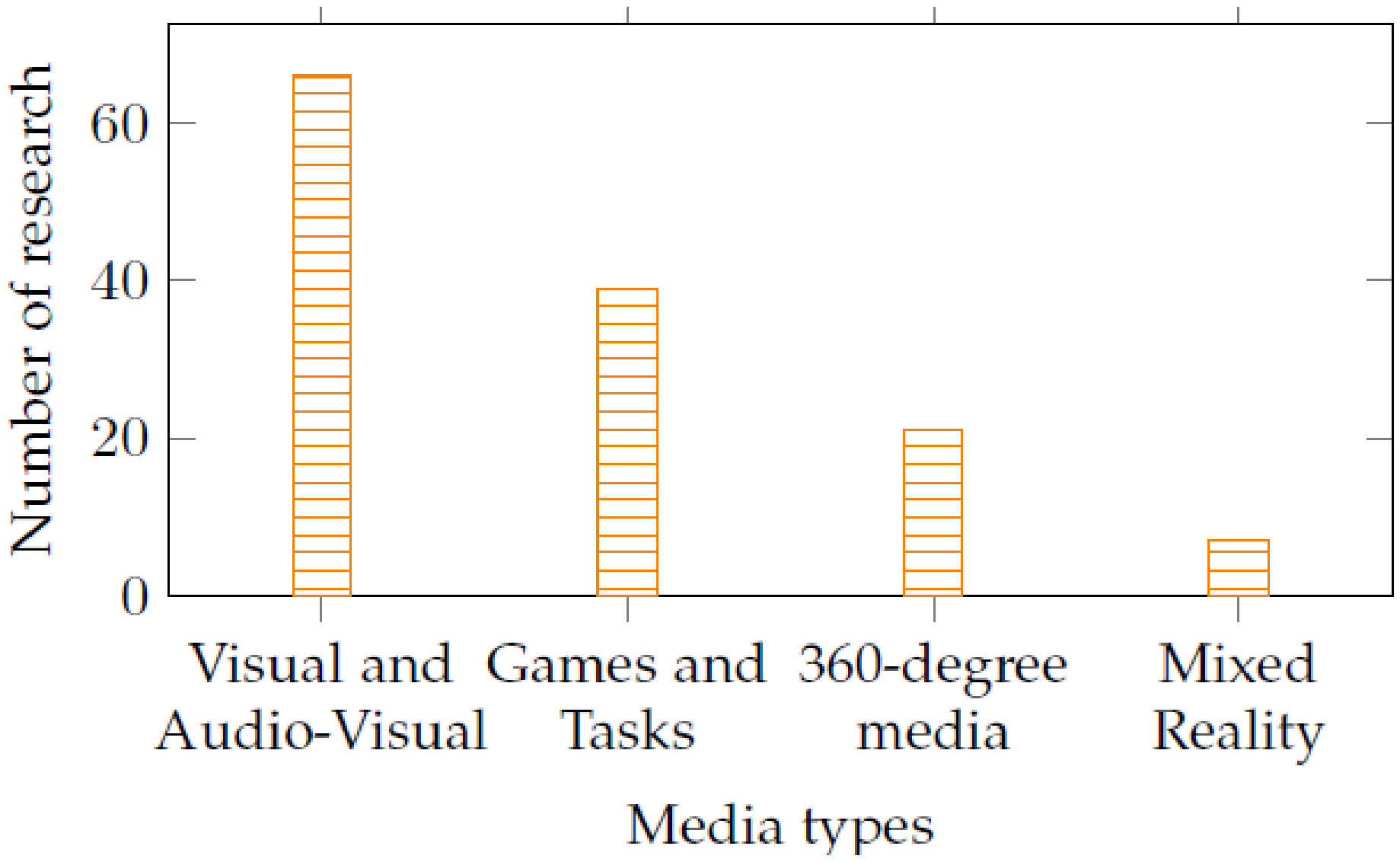

Figure 2 A graphical illustration of the number of studies that induced emotions using various VR media types.

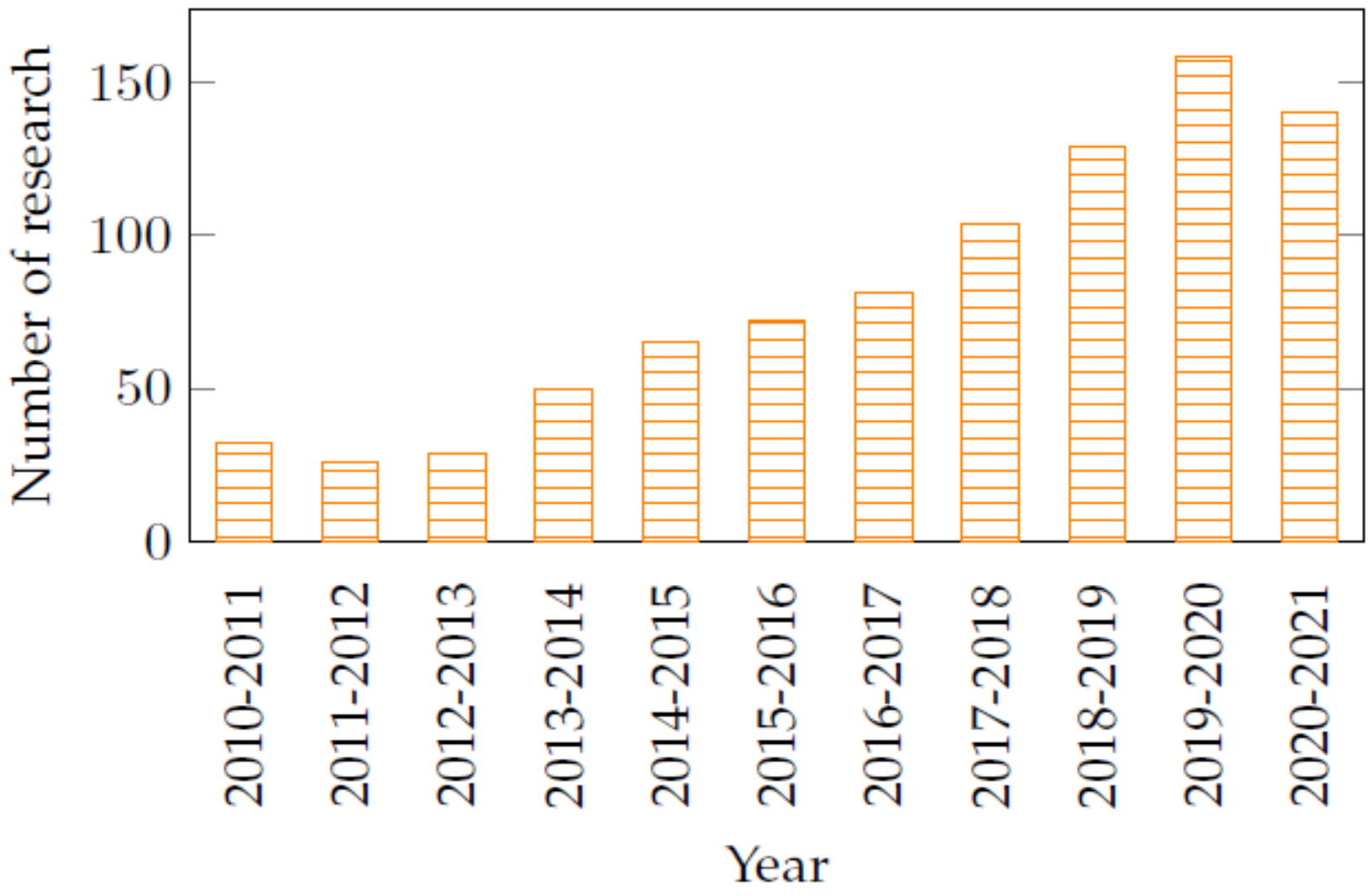

Figure 3 The graph reports the number of papers per year with the keywords "Virtual Reality" & ("Affective computing" or Emotions) & (Elicitation or Induction) obtained from Google Scholar.

The survey was constructed covering 74 papers related to VR technology in affective computing and psychology. While we have witnessed a revolutionary development of VR-related studies in the literature, we have limited the focus of this review to the evaluation of the emotional induction material, which is the primary basis for affective computing studies. As technology evolves, researchers need to study and apply those techniques in academic implementations. Consequently, validation of state-of-the-art technology is required to design an innovative and reliable paradigm. This review is among the first surveys in emotional studies, that the effect and feasibility of VR to study emotions are analysed. In addition, we investigate the potential to induce discrete and dimensional representations of emotions using several VR media contents. The results can inform the type of media content to be used to evoke emotions successfully in future emotional studies.

The rest of the paper is organized as follows. Section 2 describes the main emotional frameworks and then in section 3 on emotion elicitation, we discuss what elicitation is and the main emotion induction modalities as passive and active. In section 4, we will discuss in detail VR and report the main media content that can be used to induce emotions by VR. Next, the possibility of VR to evoke discrete and dimensional space is reviewed in section 5. In section 6, we present the public datasets generated using VR, and subjective measurements in such studies. After that we discuss the emotion sensing interfaces in section 8 and the limitations of using VR in emotional research in section 9; and, finally, we conclude by directing future research.

## 2 Emotion representation

Emotions are a major factor influencing human development. It is an aggregated sensation associated with mental, psychological, and physiological changes. Accordingly, emotional modeling and computerization are critical domains in the fields of psychology, computer science, and cognition which was later termed as Affective Computing (AC) [6]. Numerous paradigms have been developed to study emotion by data-driven and theory-driven approaches such as Discrete, Dimensional, and Appraisal frameworks.

Being one of the theoretical frameworks in AC, discrete models hypothesize that there are a significant number of core emotions each arouses a response to address an evolutional requirement [28, 29]. Grandjean, et al. [30] identified basic emotions as affect programs triggered by events that cause changes in facial expressions, physiology, and functionality. Contrariwise, Ekman [31] argued that basic emotions are common in human and species cultures, so the need to define core emotions are reasonable, while others are more likely to vary between cultures and are unique to species. Gu, et al. [32] defined basic emotions as internal regulations of neuromodulators that cause external behaviours, and as primary, inner conditions that have been adjusted through evolution to deal with the individual goals derived from the background. By contrast, Ekman [33] discussed several features that distinguish basic emotions from others, highlighting distinctive universal signals, automated assessment mechanisms, emotion-oriented physiology, and universal antecedents as prominent.

Notwithstanding the acceptance of the basic theory of emotion, there is no agreement on the exact number of core sensations [32]. For instance, after closely observing facial expressions, Ekman [31] suggested a more extensively used model that contemplates six fundamental emotions: fear, surprise, happiness, anger, disgust, and sadness. Izard [2, 34] categorized basic emotions as positive (joy, interest) and negative (fear, sadness, disgust, anger). Additionally, he emphasized that social emotions (guilt, shame, contempt) and emotional patterns of love and attachment can be conditionally incorporated into basic emotions assuming their principles for evolution, development, psychology, and adaptation.

Table 1 illustrates the discrete views of emotions defined by various theorists, and implicitly demonstrates that many theorists have agreed to define basic emotions like happy, fear, sad and anger. Apart from them, several psychologists also consider surprise and disgust as emotions within the basic framework.

Table 1 Discrete Emotions Defined by Psychologists

| Reference | Theorists | Discrete emotions |
|---|---|---|
| Izard [34] | Carroll Izard | Joy, interest, fear, sadness, disgust, anger, guilt, shame, contempt, love, attachment |
| Egger, et al. [35], Gu, et al. [32], Domínguez-Jiménez, et al. [36] | Robert Plutchik | Acceptance, anger, anticipation, disgust, joy, fear, sadness, surprise |
| Ekman [31], Gu, et al. [32] | Paul Ekman | Happiness, surprise, fear, sadness, anger, disgust |
| Ekman, et al. [37] | Paul Ekman and Wallace V. Friesen | Anger, happiness, fear, surprise, disgust, sadness |
| Parrott [38] | W. Gerrod Parrott | Anger, fear, joy, love, sadness, surprise |
| Frijda, et al. [39] | Nico Frijda | Desire, happiness, interest, surprise, wonder, sorrow |
| Jack, et al. [40] | Rachael E. Jack, Oliver G.B. Garrod and Philippe G. Schyns | Fear, anger, joy, sad |
| Gu, et al. [41] | Simeng Gu, F. Wang, T. Yuan, B. Guo, and J. H. Huang | Fear, anger, joy, sadness |
| Pereira Junior and Wang [42] | Fushun Wang and Alfredo Pereira Junior | Fear, anger, joy, sadness |
| Zheng, et al. [43] | Zheng, Simeng Gu, Yu Lei, Shanshan Lu, Wei Wang, Yang Li, and Fushun Wang | Fear, anger, joy, sadness |
| Gu, et al. [32] | Simeng Gu, Fushun Wang, Nitesh P. Patel, James A. Bourgeois and Jason H. Huang | Happiness, sadness, fear, anger |
| Cicero and Graver [44] | M. T. Cicero and M. Graver | Fear, pain, lust, pleasure |
| Grandjean, et al. [30] | Didier Grandjean, David Sander, and Klaus R. Scherer | Anger, joy, sadness, fear, disgust |

The dimensional theory of emotions states that emotional terms can be defined by a point in a continuous space of numerous continual elements [29]. From the beginning of the definition of dimensions as pleasure-unpleasantness, excitation-negation, stress-relaxation [30], these models play a major role in emotional evaluation. Following the initial interpretations, several models emerged, with variations in the number and type of scales. Among them, the two-dimensional Circumplex Model of Affect [45] is the most widely used, representing the entire affective states by valence and arousal [46]. Valence defines the range from positive to negative, and Arousal characterizes the activation of emotions ranging from active to passive [35]. In addition to these two dimensions, another scale was introduced as the Dominance to the PAD model to represent control [23]. Nevertheless, the following Table 2, which reviews the dimensions hypothesized by various psychologists, shows that they have come to a weightier conclusion for interpreting emotion dimensions as valence and arousal as shown in Figure 4.

Table 2 Emotion Dimensions Defined by Psychologists

| Reference | Theorists | Dimensions |
|---|---|---|
| Schlosberg [47] | Harold H. Schlosberg | Pleasant-unpleasant, tension-relaxation, excitation-calm |
| Ekman [48] | Paul Ekman | Pleasant-unpleasant, active-passive |
| Russell [45], Posner, et al. [46] | James Russell | Valence, arousal |
| Wundt [49] | Wilhelm Wundt | Valence, arousal, intensity |
| Osgood, et al. [50] | Charles Egerton Osgood, William H. May, and Murray Samuel Miron | Arousal, valence, potency |
| Grandjean, et al. [30] | Didier Grandjean, David Sander, and Klaus Scherer | Valence, arousal |

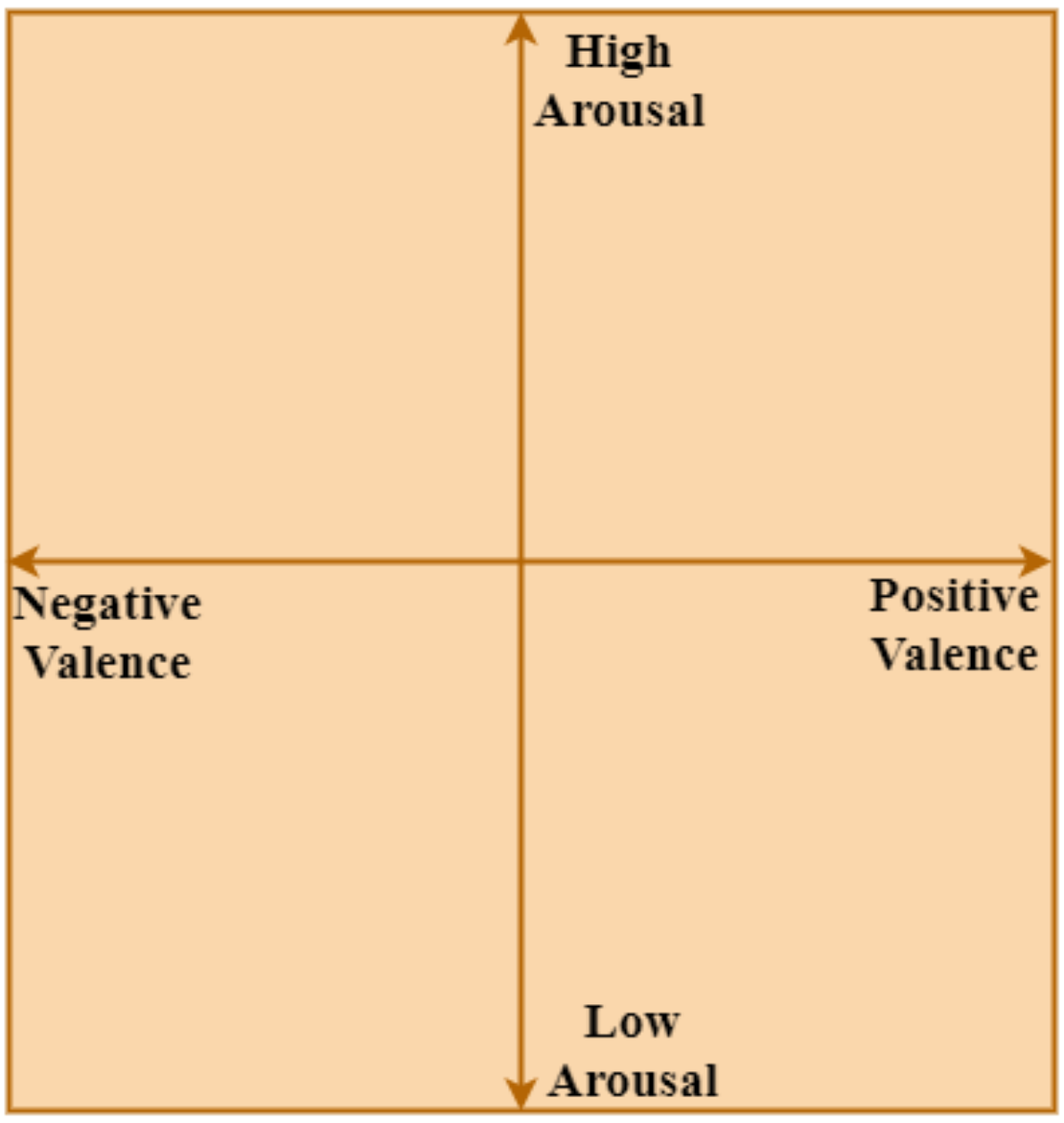

Figure 4 A graphical representation of the two-dimensional models of emotions with valence-arousal axes.

The other emotion framework which has recently gained more attention in the AC community is the appraisal model. Several studies have already highlighted the importance of taking a multi-componential approach to emotions [18, 51]. Appraisal theories address emotions as processes and components [52]. The main components identified by appraisal theory on the emotional episode can be identified as; an appraisal component which evaluates the event, a motivation component that defines action tendencies, a physiology component with changes in bodily functions, an expression component with motor expression and a feeling component showing emotional experience. Component Process Model (CPM) which is a variant of appraisal theories has four major evaluations in the appraisal component as; relevance (how the event is related?), implications (what are the implications of the event?), coping potential (how to cope with the event?), and normative significance (what is the significance of the event?) [53]. However, yet there are very limited studies available in the literature that assume the appraisal theories, even though these models describe the full emotional experience rather than a single perspective like feeling. Therefore, in this review, we will not consider appraisal theories in detail. However, we recognized the importance of considering the full appraisal components in AC, and the necessity to go beyond the two dimensions (valence, arousal) to define the emotional experience [18, 51].

## 3 Emotion elicitation

Induction mechanisms play a crucial role in triggering emotions by simulating the sensations required for a research plan [54]. In human affective disciplines, it is essential to elicit affective states that are decisively trustworthy and morally influential [55]. Choosing the right elicitation mechanism is essential for the success and quality of research behaviour and for drawing reliable conclusions.

Stimulus presentations should be organized in a manner that provides the same viewing experience for all participants [16]. Previous implementations seek to control for the stimuli presenting duration [22, 56], screen resolution [16, 57], screen brightness [16], screen size [26], VR headset types, environmental conditions [58], standing or seating position. Also, most of the studies organized all the participants to experience the same content in a randomized manner [18, 21]. This provides the generalizability of the research outcome eliminating any biases due to the content order or visual characteristics.

The evaluation of studies, performed thus far, confirm the use of different emotion elicitation sources in controlled experiments. Studies in the literature have adopted mainly two methods for emotion induction as passive and active [16]. Passive elicitation involves the subject as an observer of an emotional event and, conversely, in an active scenario subject is actively participating in the emotional encounter. The following sections will therefore focus on passive and active emotion elicitation modalities.

### 3.1 Passive elicitation

Most of the studies over the years have been researched by a passive elicitation mechanism and these include watching images (International Affective Picture System (IAPS); [59], Geneva Affective PicturE Database (GAPED); [15]), observing emotional video [12, 13, 51], recalling experience [24], listening to music (International Affective Digitized Sounds (IADS-E); [60]). For an exemplar, AlzeerAlhouseini, et al. [61] have used picture clips from the IAPS to generate happiness, calm, fear, and sadness, and showed the relationship of Electrocardiography (ECG) with arousal and higher reliability of Electroencephalography (EEG) than ECG in identifying emotions. Similarly, Katsigiannis and Ramzan [62] have used film clips to evoke nine emotions and have achieved accuracies higher than 61% for each of valence, arousal, and dominance. As another paradigm, Estupiñán, et al. [22] manipulated the images of GAPED in an immersive VR environment to evaluate the possibility of VR to improve the emotional experience by images. They identified an increase of arousal and a higher negative experience of valence in VR conditions.

A recent survey by Siedlecka and Denson [24] presented a qualitative review of five emotional stimuli assessed on visuals (image, video), music, recall, situational procedures, and imagery review against anger, surprise, fear, disgust, and sadness. They found that visual stimuli were the most powerful model for the emotional stimulation of six selected emotions. Moreover, Domínguez-Jiménez, et al. [36] used video clips to trigger happy, amusement, and neutral emotions. They have reported an accuracy of 100% in predicting target emotions. On another note, the use of films as a mode in AC comes with the desirable features as being dynamic, easy to instal, accessible, and ecological valid [63]. There is a wealth of literature on the effectiveness of video stimuli in creating a multiplicity of emotions [3, 36, 64, 65].

Broadly speaking, extensive use of images in the literature can be justified by its simplicity [35], cost-effectiveness [23], flexible processing, and easy control of experimental features [16]. Contrariwise, the lack of environmental validity [18] has led to criticism of these traditional approaches in assessing the strength of the impact. Also, observing emotional material limits the full-dimensional experience with a real sense of emotional variations [66]. On another note, static images are influenced by environmental factors and inadequate emotional immersion [67].

Although passive mechanisms are used extensively due to their readily available characteristics, they potentially limit the generation of reliable theoretical frameworks due to the observational role of the subject in the induction process which lacks any direct self-relevance. The following section will therefore focus on active participation.

### 3.2 Active elicitation

The quality of understanding the emotional process depends on the naturalness of the sensations [68, 69], social acceptability and believability [11]. Accordingly, active emotional stimuli are an optimal solution that led to the generation of more intense experiences. The presence of genuine interactive content in a shared space facilitates natural social interactions and emotions [70]. Active methods mainly involve: 1) interactions with humans and computer-generated avatars that respond with facial expressions and behavioural changes, [71-73], 2) virtual reality [20, 74] which provides the viewer with an immersive and interactive experience, as in the realistic world, and 3) games that provoke user engagement while completing tasks or tackling challenges [75, 76].

In contrast to passive methods, active methods are high in ecological validity [16] and immersivity in an interactive setting [77]. However, most studies in the literature have focused on passive methods. Possible reasons for this may be that there is a great wealth of literature, which provides standards, norms, and practices of experimental conduct using readily available passive content to induce emotions. Therefore, the selection and evaluation of emotional content can be done by referring to the literature [3, 51, 65]. To our knowledge, however, there is no literature-informed analysis of VR equipment usage in emotion studies, their type of media contents, screening processes, and the public datasets. Hence, it emphasizes the need for a literature analysis and a guide to VR-based emotional research. Therefore, the subsequent assessments will focus on those areas to fill the gap.

## 3.3 Virtual Reality

Virtual Reality (VR) is a computer-aided design that provides an interactable virtual three-dimensional environment [78, 79]. VR has been used as a medium for games [20, 80], entertainment [58, 81, 82], education [78, 83], physical and cognitive training [84, 85], rehabilitation and therapy [21, 58, 86-88], and surgery [89], as VR promotes user connectivity and naturalness. Recent studies have revealed that there is a significant increase in interest in human emotional behavioural studies using VR to effectively evoke emotions in the data collection procedure. VR can be integrated with numerous elicitation modes as images [22, 90], videos [56, 91], games [18, 19], and 360-degree panoramas [92-94]. Lists of literature on VR usage in Affective Computing, psychology, and Human-Computer Interaction (HCI) based on the content type is provided in the Supplementary material for interested readers; visual and audio-visual (Supplementary material A), games and tasks (Supplementary material B), and 360-degree panoramas and videos (Supplementary material C).

VR sources have the power to induce emotions in dimensional space. For example, Marín-Morales, et al. [92] successfully developed a computer-based emotion predictive model by eliciting active emotions via Immersive Virtual Environment (IVE). Their finding validated the use of VR in eliciting and recognizing emotions. Another similar study based on Virtual Reality (VR) games was able to achieve satisfactory accuracy in classifying emotion for valence and arousal dimensions [19]. The reader may refer to the following works which had investigated arousal [95-99], valence [100, 101], valence and arousal [19, 56, 85, 92, 102, 103], and valence, arousal, and dominance [68, 70, 83, 104, 105] using VR.

Earlier research has used VR as a medium to induce discrete representation of emotions. For example, Felnhofer, et al. [106] have elicited joy, sadness, anger, anxiety, boredom by five virtual park scenarios. Also, Meuleman and Rudrauf [18] have used VR games to elicit a diverse range of discrete emotions and showed the potential of VR in AC. Interested readers can refer to following literature for more details, that have investigated discrete representation of emotions using VR [20, 21, 25, 26, 56, 58, 70, 75, 77, 80, 82, 95, 104, 105, 107-115].

The success of VR can be explained by its ability to perform better than traditional non-immersive content and to keep subjects immersed in a collaborative environment [116]. Clearly, with the usage of low-cost Head Mount Display (HMD) [74], VR technology has proven to produce immersion [20, 26, 77, 117], a sense of presence [20, 66, 83], interactivity [110], and isolation of participants from external stimuli [92] in controlled experiments. For example, Hidaka, et al. [66] recorded VR scenes of happiness, depression, relaxation, fear, and distress which were viewed by the subjects. They concluded that the VR sources provided by the wearable HMD were more effective than the traditional display where they achieved a higher average value for effectiveness, efficiency, and environment setting for HMD than the display.

A VR system usually consists of a headset with visual and audio facilities, controllers, and tracking stations (or tracking built-in using computer vision approaches for localisation). As given in supplementary material A, B and C, recent research mainly uses HTC Vive[1] and Oculus Rift[2] HMD for their research. One reason may be that VR materials such as games, videos, and scenes, are readily available on streaming platforms as Vive Port[3] and Steam[4] for HTC Vive and Oculus Store[5] for Oculus Rift. HTC Vive platform is affordable [74], provides free movements and accurate positioning within the boundaries of virtual space [118], and can be integrated with VR developer engines [118-121]. Similarly, Oculus Rift is available at an affordable range [78, 112, 118, 122], is portable [78, 122], can be easily integrated into game engines for software development [78], and provides precise location details in the VR space [78]. Moreover, both of these commonly used HMDs are designed as wired, wireless, consumer graded [75] and provide 360-degree rotations [58, 91, 123].

The novelty of VR for emotion studies arises from its' ability in triggering ecologically valid immersed emotive content in controlled research settings [69, 92]. VR can immerse participants in the virtual world and users can get fully engaged in that narration with intense emotions [20, 124]. It has the potential to effectively elicit emotional events leading to synchronous changes throughout the participant's entire body [18, 20]. Therefore, VR can be an optimal solution to date for the study of emotions to evoke natural feelings.

# 4 Elicitation material related to Virtual Reality

## 4.1 Visual and Audio-Visual Stimuli

Visual stimuli can be static media content such as images (e.g., static scenes), or dynamic content such as movies and Virtual Environments (VE). Digital or printed images are one of the main stimulants used in academic implementations to evoke emotions in the data collection process so far [25]. Typically, researchers select images from a large dataset after assessing their ability to elicit the required emotions and present one at a time. To minimize the risk of emotional cumulation, researchers usually present specific emotions in a single image stimulus [22, 90]. In addition to the elements mentioned in Section 3 (duration, screen resolution, screen brightness, screen size, VR headset type, content), researchers usually keep the image size constant during the presentation.

For the VR-based emotion induction, researches often use colored images from the International Affective Picture System (IAPS) which has a range of events and objects that are related to humans [59] (used in [57]). These images are pre-rated on the scales of arousal, pleasure, and dominance. IAPS is widely used in several research areas to study emotions [125, 126], and medical disorders [57]. Another image dataset is the Geneva Affective PicturE Database (GAPED) [15] (used in [22]), which consist of 730 images pre-annotated with valence and arousal scores. This dataset consists of images that can be categorized as positive, neutral and four negative contents (siders, snakes, violation of legal and moral norms). Also, Nencki Affective Picture System (NAPS) [54] (used in [127]) is a publicly available database containing 1356 high-quality photos of people, faces, animals, objects and scenes. Images are provided with valence, arousal, and approach-avoidance scores as well as luminance, contrast, and entropy ratings. Nevertheless, some studies suggest combining visual content with audio content to increase the effectiveness of mood induction [106].

[1] https://www.vive.com/
[2] https://www.oculus.com/
[3] https://www.viveport.com/
[4] https://store.steampowered.com/
[5] https://www.oculus.com/experiences/quest/

Selecting static images for a study is easy; also, presenting and collecting data from participants is simple. Similar to the monitor presentation, image stimuli can easily render to the VR environment and present to the participant. However, the emotional intensity evoked by any image is low compared to the other mechanisms (video, virtual environments, games) discussed in the following sections [16]. Therefore, the intensity may not be strong enough to learn from images. While watching images, the emotional responses tend to fade quickly [16]. Also, as images are only a symbolic representation of passive stimuli, differences in physiological signals, and expressions are small. Therefore, responses to presented images are primarily collected as subjective reports in most studies. Furthermore, image stimuli are not much accountable for evoking discrete representations of emotions in comparison to dimensional representations of emotions [16].
Film stimuli are one of the common dynamic audio-visual stimuli used in the study of emotions [25] where participants follow a narrative story. The film-based studies performed thus far are mainly aimed at obtaining audio-visual stimuli through film libraries [97], film excerpts [56] and Immersive Virtual Environment (VE) [25, 128]. A neutral stimulus or wash out clip is presented before each emotional clip to generate the baseline. For neutral stimulants, researchers usually use screensavers [16], or grey backgrounds [97] as wash out clips.

Immersive Virtual Environment (VE) allows the audience to explore the entire interactive graphical world with or without control. As a distinguishing feature, VE provides researchers with the ability to simulate realistic, imaginative, and physically impossible moments and tasks. For example, Cebeci, et al. [82], collected data from subjects that allowed them to experience unpleasant emotions via a horror environment, cybersickness by a roller coaster experience, and neutral feelings by a campfire environment. Apart from VEs, scholars have implemented avatars to evoke basic emotions that demonstrate facial expressions. As an illustration, Bekele, et al. [57] have expressed happy, surprise, contempt, fear, disgust, sad, and anger at four arousal levels as low, medium, high, and extreme by computer-generated avatars playing facial expressions. Another similar research was done by Gutiérrez-Maldonado, et al. [21], where they induced happy, sad, fear, anger, disgust, and neutral emotions via male and female avatars showing changes in the facial area.

### 4.2 Games and Tasks

Gaming has ranked itself as one decisive landmark of entertainment, which has seen incredible growth over the last decade. Games can arouse a broader spectrum of affluent and diverse emotions by the collaboration between players and player-agents through effective game events and gameplay [129]. VR gamification is responsible for creation of engagement [19, 85], entertainment [19] and involvement [130]. This flow (involvement) has been studied by Granato, et al. [19] with VR racing games and analysing physiological signals. Game sessions reveal that players were enthusiastically absorbed in the elicitation session. Another research [130], studied the correlation between flow and presence, with role-playing games. They pointed out that games are a great source for getting a flow experience.

Although there are similarities between the content of the games and the audio-visual stimuli (please refer to section 4.1), there is a significant difference in the participants' level of interaction [80]; while games engage with players actively, audio-visuals engage with the audience passively. In games, players have control over the events and decision-making power, thus the next gameplay and outcomes will be based on the actions and decisions.

In terms of measurements, games can create measurable variations in physiology to objectively monitor emotional components. This was demonstrated in [131], where they used multimodal analytics: Blood Volume Pulse (BVP) and Skin Conductance (SC) to predict emotions. They observed that the proposed approach could accurately predict anxiety, fun, excitement, and relaxation using features extracted from statistical analysis (average, standard deviation, max, min, range etc.) and Convolution Neural Networks (CNN). Furthermore, the authors have used an automated feature selection where they fed selected features into training models. The results revealed an increase in accuracy while fusing both BVP and SC. In another case [85], a study of cognitive training used VR gamification tasks and collected facial Electromyography (f-EMG) signals. Their results were significant in accurately predicting emotions in terms of valence and arousal. Moreover, this result shows the possibility to use games as an appropriate stimulus to study facial expressions. Similarly, VR games, evaluated by self-reporting and bio-signals, indicated the experience of higher levels of happiness, surprise and presence [109].

In addition, games can create motivating tendencies to win, immerse, and socialize [85]. These features of games indicate the ability to effectively access the motivational aspects of emotions owing to the active involvement of the subjects. Most importantly, due to the rapidly changing nature of a game, we can expect variability in the emotions of the subjects [19]. This is a good alternative for studies focusing on models based on appraisal theory [13, 51] that assume rapidly changing processes [132].

With the advent of VR technologies in the gaming market, the use of traditional video games has changed dramatically to an immersive state [75]. In VR, the player actively encounters events like the real environment. VR controllers allow players to interfere with actions, tasks, and touch in a mediated environment similar to the real world. Because of that VR games are more advantageous than traditional video games [80, 109].

Further, games have a challenge that evaluates a player in terms of their physical and mental involvement [20, 83]. In the physical challenge, the player is evaluated mainly through accuracy, speed, strength, coordination and in the mental challenge player is evaluated through cognitive effort, reasoning, decision making, inspecting, and planning [20]. Peng, et al. [20] conducted a study with several VR games with or without emotional challenges such as reasoning, decision making while the family is facing a sudden nuclear war. They found that games with emotional challenges had the potential to provoke different emotions than traditional challenges and emotions are more enriched in VR. Moreover, they concluded that an emotionally challenging game environment with higher vagueness, virtual characters, complex topics, and actions are more capable of revealing a naturalistic emotional experience.

According to Meuleman and Rudrauf [18], even though non-VR video games involve participants actively, they still separate the subject from the game narration by an avatar. So typical video games still lack in providing complete immersion and subjective feeling.

Therefore, VR games are a better medium for emotional studies to provide a fully dimensional experience [18]. The use of VR games as an active eco-valid method in emotion-related studies can be highlighted as successful because of its ability to perform better than traditional content and to immerse subjects in a collaborative environment. Researchers usually select games from gaming platforms like Steam[6]and Oculus[7]. List of VR games used in literature with dominant emotional experience is given in the Supplementary material D.

### 4.3 360-degree panoramas and videos

This is a technology that provides a 360-degree panoramic view of a virtual environment. By integrating to the VR technology, it provides the ability to view the background or scene in a full rotation. For example, Marín-Morales, et al. [92] generated architectural environments[8] that can be visualized in 360 degrees. They generated four environments that are relevant to valence-arousal space by modifying the factors such as illumination, colour, and geometry. Using Support Vector Machine (SVM) they achieved an accuracy of 75% for arousal and 71.21% for valence prediction. Also, [133, 134] yielded significant results for the four-class classification of emotions which were induced via 360-degree videos based on Russell's model of emotions. Further, a 360-degree video dataset was established by Suhaimi, et al. [77] using commercially available video content from streaming platforms. Those videos were based on a couple in a relationship, skydive experience, evil encounter, mysterious underwater experience, golden retriever puppies, bunnies, tree climbing experience, and an experience from a tower.

## 5 Different emotions in Virtual Reality

This section analyses the possibility of using diverse VR media types for the induction of different types of emotions. Supplementary material E provides an overview of the research works done using VR, including the details about stimuli content, outcome, annotations, and measures. Accordingly, the increasing number of publications with a successful outcome, indicates that VR has a great potential in studying emotions, and triggering changes in physiology.

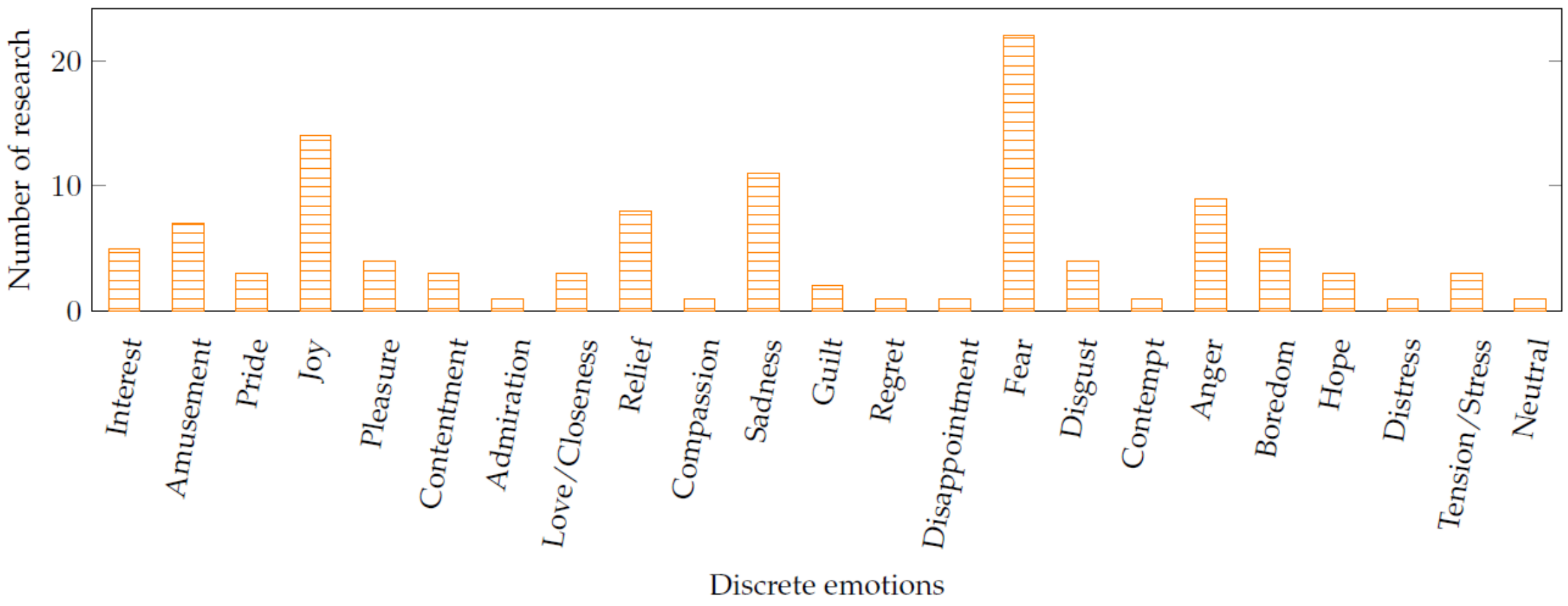

Figure 5 Distribution of discrete emotions induced in the VR literature.

[6] https://store.steampowered.com/steamvr
[7] https://www.oculus.com/experiences/rift
[8] http://personales.upv.es/jamarmo/emotionalrooms/

Figure 5 shows the distribution of main discrete emotions studied by the publications given in Supplementary material E. According to the illustration, it can be noted that VR material has been frequently used for inducing fear, joy, sadness, anger, relief, and amusement. However, emotions like pride, contentment, love, hope, tension, guilt, admiration, compassion, regret, disappointment, contempt, and distress have not been reported frequently in the literature. This may be due to the high complexity of those emotions, inattention of affective studies to study those emotions via VR or lack of readily available VR contents to induce such emotions. Given the higher efficacy of VR in generating a realistic experimental paradigm, future research should investigate ways to induce such emotions more effectively.

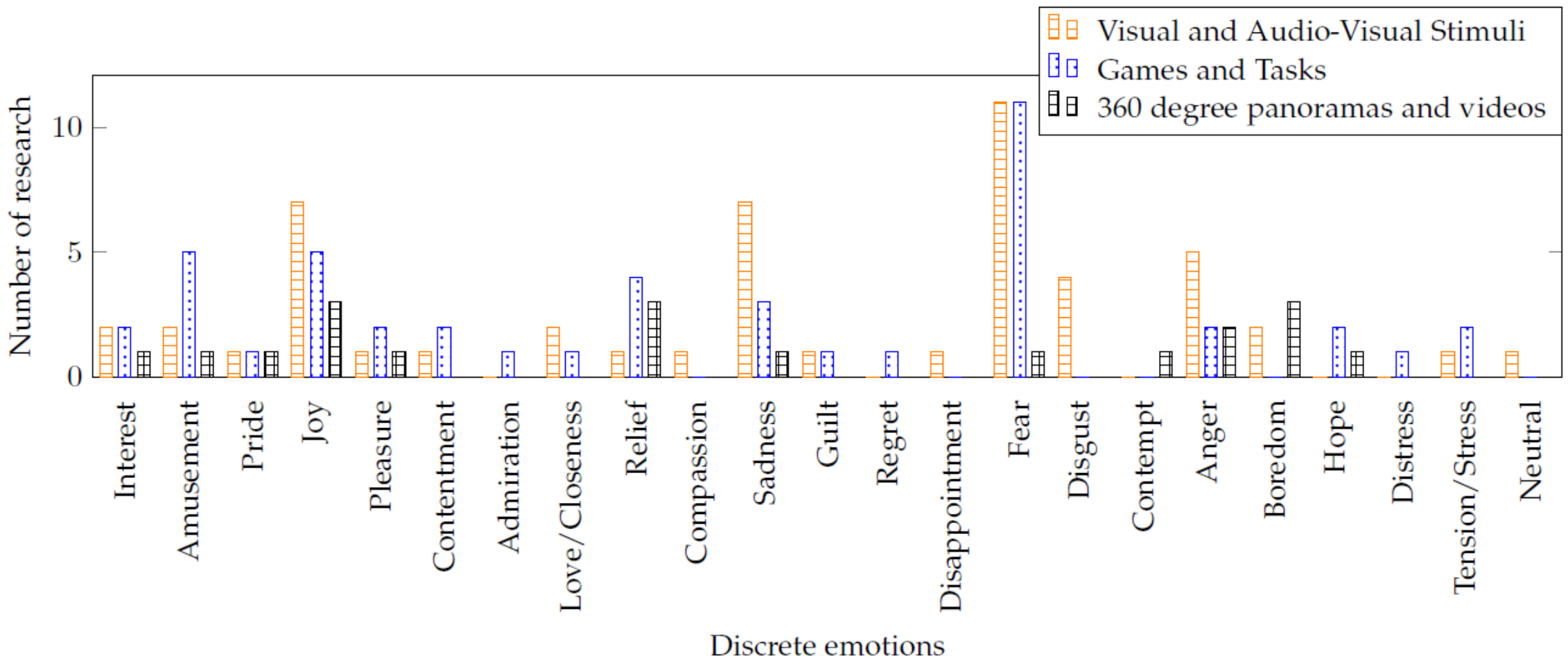

Figure 6 Frequency of discrete representation of emotions experiences by visual and audio-visual stimuli, games, and tasks, and 360-degree media in VR.

Figure 6 shows the frequency of each emotion experienced by research publications based on the material used. Accordingly, it can be observed that visual and audio-visual stimuli seem to be used more often to induce happiness, sadness, and anger compared to the rest of the material. This can be explained by the minimal availability of the materials that induce diverse emotions and the less attention in the domain to utilize and generate relevant VR content. Games and tasks have been more widely used in inducing amusement, pleasure, contentment, relief, fear, and tension than the other two types. And it has been least used in pride, admiration, love, guilt, regret, distress, compassion, disappointment, disgust, contempt, and bored. This bias towards using games for the positive emotions is not very surprising as it would be challenging to find games that activate negative emotions as game developers focus primarily on entertaining the community. However, 360-degree media doesn't seem to be used very often compared to visual and audio-visual stimuli and games. Altogether, the possibility of using VR as an affective medium to study a wide range of emotions is validated. However, further research works are required to arrive at more solid conclusions related to complex emotions.

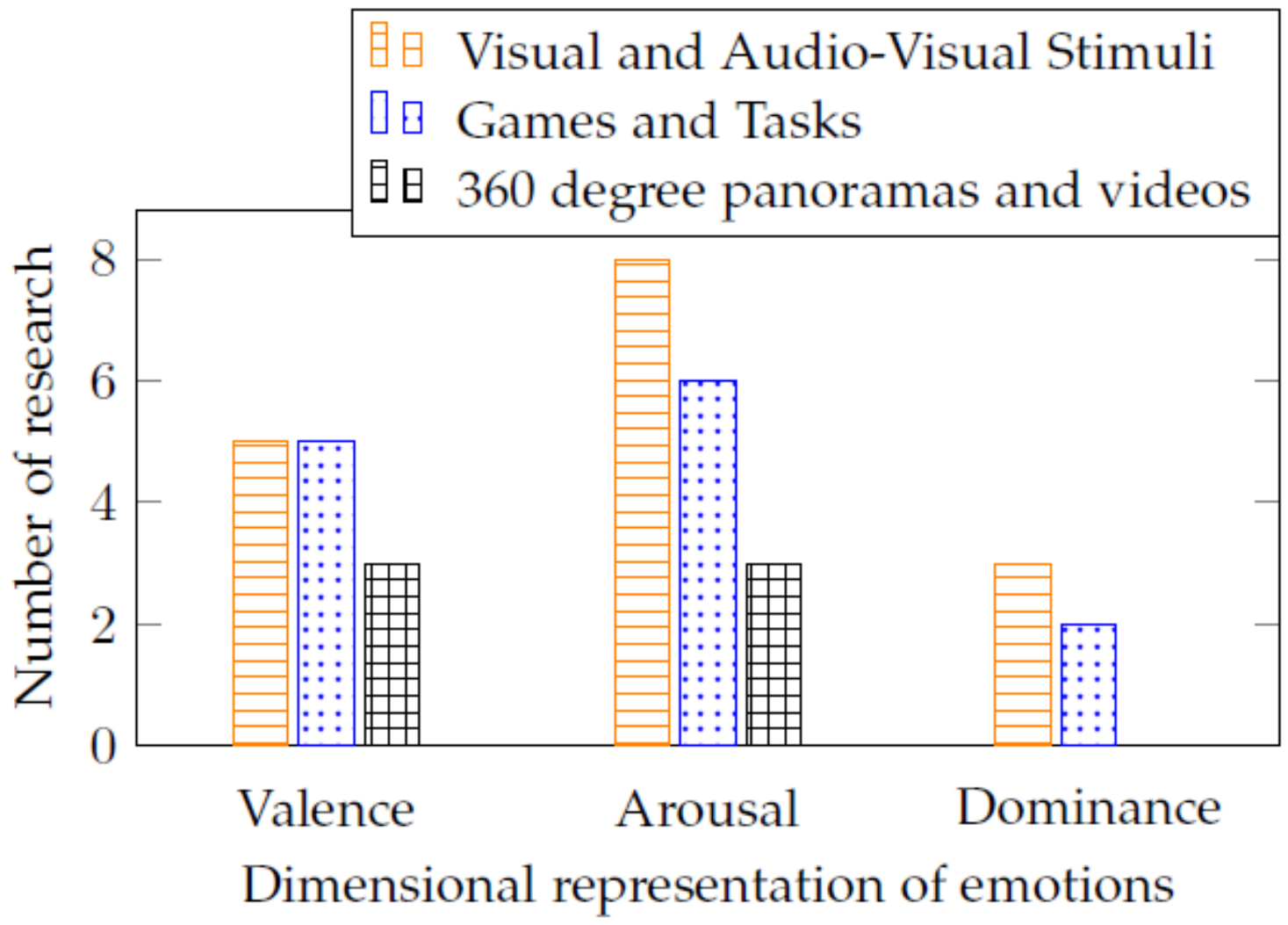

Figure 7 Frequency of dimensional representation of emotions by visual and audio-visual stimuli, games, and tasks, and 360-degree media.

Figure 7 shows a summary of research works conducted on dimensional model of emotion using different materials. Thus, within the scope of our selection criteria, visual and audio-visual stimuli, games, and tasks have become an effective medium for the study of valence, arousal, and dominance and have been frequently used for prior research. However, there is limited research to conclude about dominance. Overall, similar to before, 360-degree media has been used less frequently; however audio-visual content and games have been used more commonly to elicit valence and arousal.

## 6 Datasets

Generation of an empirical database on affective computing is quite difficult owing to the lack of necessary expertise [135-137], unavailability of gold standard equipment [135], the involvement of subjects, time constraints, lack of budget, proper controllable environment and portable devices [62, 64]. Therefore, public datasets are a great resource for scholars to augment research on emotion, mood, and feeling perspectives [62, 135, 138]. Hence, in this section, we will review publicly available datasets generated using VR.

### 6.1 Datasets created using video

Recognizing the contemporary academic gap without a rich VR media database, the authors of the work [91] generated a public dataset for scholars. This dataset contains 73 immersive VR video clips which are available online[9]. Each video was annotated with a valence and arousal rating, duration information and with a description.

### 6.2 Datasets created using games and tasks

The dataset collected by Caldas, et al. [83] is available online[10] and has a total of 87 recordings from the participants who played five trials of rehabilitation-based VR game (a skydiving game). This dataset includes demographic information of the participants, experienced emotions after each game play (valence, arousal, and dominance) and physiological signals (electrocardiogram, skin conductance and respiration) collected during each game play.

[9] https://vhil.stanford.edu/360-video-database/
[10] http://dx.doi.org/10.21227/vj8w-v224

Delvigne, et al. [139] created a database[11] of EEG and eye-tracking signals to study the state of attention. Data were collected while 32 participants were completing tasks in five everyday life VR environments (bedroom, gym, park, living area, forest). This database can be used in the areas of participant recognition and attention assessment. On another note, processed data from the study by de-Juan-Ripoll, et al. [119] with 98 participants are available online[12] . Their VR environment was based on two mazes where participants need to complete individually and with virtual avatars. Risk is spread in the maze by bridges, animals, storms, and haunted areas and players can avoid risks by shields. Their study aimed to assess participants' personality, sensation seeking, behaviour, and biosignals (Galvanic Skin Responses) in risk environments.

### 6.3 Datasets created using 360-degree panoramas and videos

Being a comparatively innovative field of research in emotional studies, one public dataset is available online[13] which was generated by Marín-Morales, et al. [92]. This data source consists of 360-degree panoramas of architectural buildings that cater for dimensional model of emotion. Each stimulus is annotated with valence-arousal tagging. Another similar data source[14] has been made public, where they elicited emotions through 62 360-degree videos[15] [103] that were used in previous research [91]. Their dataset is illustrated by valence-arousal cataloguing using Self-Assessment Manikin (SAM) [140] and EmojiGrid affective scales.

## 7 Measuring subjective experience in Virtual Reality

To investigate research objectives and motivations, most VR-based studies apply a subjective experience evaluation, in which participants are asked to imply about the experience through some standard questionnaires. These self-reporting questionnaires are presented in paper-based, digital [97, 105], verbal [56, 141] and VR [19, 103]. In emotional studies, researchers often focus on the evaluation of discrete [80], dimensional [19, 91], and appraisal [18] models of emotion. To investigate discrete emotional components, researchers tend to define the self-reporting criteria based on emotional frameworks such as Geneva Emotion Wheel (GEW) [142] (used in [18]), Positive and Negative Affect Schedule (PANAS) [143] (used in [21, 69, 104]), Differential Emotions Scale (DES) [144] (used in [106]) Emotion Annotation and Representation Language (EARL) [145] (used in [20]), Visual Analogue Scale (VAS) [146] (used in [25, 82]) and verifying the suitability with selected stimuli. For the evaluation of dimensionality of emotions mainly arousal, valence, and dominance are considered; and most studies have used one or a combination of available assessment scales such as Self-Assessment Manikin (SAM) [140] (used in [67, 91, 92, 95]), Affective Slider (AS) [147] (used in [19]), and Russell's circumplex model [45] (used in [85, 122, 148]). Apart from that, novel immersive self-reporting tools are available for assessing valence-arousal space such as EmojiGrid [103]. Appraisal theories have mainly been evaluated using a variation of GRID questionnaires [149, 150] (used in [18]).

[11] https://vdelv.github.io/dataset.html
[12] https://github.com/ASAPLableni/AEMIN-Dataset
[13] http://personales.upv.es/jamarmo/emotionalrooms/
[14] https://osf.io/482ne/
[15] https://www.stanfordvr.com/360-video-database/

## 8 Emotion sensing interfaces for Virtual Reality

Emotions transmitted can be assessed through a variety of bodily indicators using different technologies, methods, and tools. These approaches differ from each other by their performance in emotion differentiation, accuracy, usability [35]. The study of emotion involves the development of various frameworks for effectively automating emotion detection. Most such models have laid the foundation based on the analysis of facial, speech, gesture, and physical phenomena [151]. Nevertheless, the VR headset covers about half of the user's face, preventing face expressions from being detected by traditional camera tracking methods. Therefore, embedding physiological sensors to the user's face and collecting data is highly accepted in contemporary studies.

There is a wealth of literature on the use of wearable technologies for AC. Picard and Healey [152] hypothesized the concept of recognizing affective states of a wearer equipped with sensors and devices and termed as "Affective Wearables". Prolonged physical connection with the user helps to understand the individual patterns of human behaviour associated with the affected states. Over the past few decades, extensive research has been conducted to develop emotion recognition applications with several biological senses. As suggested by AC concepts [6], the system layout, which includes physiological signals, is more advantageous than conventional methods such as self-reporting, expert reviews due to the inability to hide and control [36, 61, 153] as they are unconscious responses that occur after an event [9]. One of the positives of using biosignals is the ability to create wearable and seamless systems [36]. Furthermore, they are responsible for the continuous measurement of signals throughout the signal acquisition process [66].

The continuous evolving of VR based affect recognition technologies have formulated a novel research trend to incorporate wearable physiological sensors into VR headsets in response to the less portability and mobility of conventional biometric sensors. Recognizing that, real-time data acquisition interfaces have been developed and used in some of the recent literature [85, 116, 154]. So, the following will focus on the contemporary emotion-sensing interfaces for VR and their applications.

Facial Electromyography (f-EMG) is a technique to evaluate facial muscle functions through electrical pulses. But, in the context of VR, computer vision analysis of facial expressions is not appropriate as the user's face is largely covered by the headset. Hence, among the many physiological signals, EMG-based methods can be used as a more reliable approach to the study of facial expression in VR due to the correlation of muscle activity with emotions [155]. Previous research has shown the potential for emotional recognition through built-in [19, 155, 156] and wearable equipment [85, 116, 157], and commercial versions of the f-EMG interface are currently available, considering the possibility of using f-EMG as a means of emotion recognition due to the onset of feelings of muscle activity.

EmteqVR[16] system [116], which is the latest progression of Faceteq [158] is commercially for biometric analysis of a VR immersive experience. This platform consists of dry electrodes that are embedded into the emteqGo VR headset or as an interface that can be embedded into the HTC Vive Pro headset [159]. This device comes with seven-channel f-EMG electrodes, a Photoplethysmography (PPG) sensor, an Inertial Measurement Unit (IMU) and real-time data manipulating software. Moreover, studies such as the one conducted by Reidy, et al. [85], demonstrated that EMG signals from Faceteq HMD can be used to differentiate valence and arousal with an accuracy of 64% and 76% respectively. In another study using the Faceteq interface, researchers identified arousal detection from PPG sensors [97].

[16] https://www.emteqlabs.com/

Intending to improve the flexibility of using plug-and-play interfaces with VR HMD, researchers have developed PhysioHMD [154] which is a software and hardware interface for collecting physiological signals. The interface has an array of EEG, EMG, EOG, EDA sensors that can be connected to AR or VR headsets. Their system is integrated into the Unity3D package so that the end-user can easily configure the developer environment and use visualizations for behavioural research.

Another potential interface that can be used in VR research for affective science was introduced by Looxid Labs[17]; LooxidVR headset as a complete VR system and LooxidLink as a VR mask; [160]. The LooxidVR headset enables researchers to integrate to Unity, time sync, track eye movements and EEG signals (6-channel). LooxidLink which is a non-invasive mask can be integrated into Oculus Rift, HTC Vive and Pro versions track 6-channel EEG [161].

Researchers have also explored wearable biosignal sensing devices that can be integrated with commercial VR headsets. For example, [120] has used a 14-channel Emotiv EPOC+[18] EEG headset in VR based research, implicitly illustrating the feasibility of a unified experimental design. Similarly, Neurable[19] has developed a retrofitted EEG headset that can be assembled into a VR headset and provides a natural translation of user intentions [162]. Moreover, a commercialized wearable VR is manufactured by OpenBCI[20]. Lakhan, et al. [163] has evaluated the possibility of using OpenBCI EEG device in comparison to research-grade EEG systems for emotion recognition and has claimed the significant accuracy of the device. Researchers have been interested in incorporating this wearable device for neurological research formulated by a VR protocol [81, 101]. Muse[21] is another wearable that has been used in VR-based studies [98, 99, 115, 164] to obtain EEG signals.

With the perception of VR-embedded biosignal sensors in mind, these emerging emotion sensor interfaces could expose modern directions for affect and activity recognition, heading to potentially improved knowledge of human behaviour. These technologies, however, have little evidence in neuroscience and psychology research, hence, require further attention and improvement to develop a broader research domain.

## 9 Practical Implications in using Virtual Reality

While VR has been increasingly used in the research domain, there appear to be a set of norms and standards to be followed in practical use. Employing these customs may greatly elevate the efficacy of research conduct and minimize the impact of the limitations. First, researchers should evaluate the probability of different kinetic environments on the average population to lessen the impact of motion sickness. This facilitates in filtering participants at an early stage to collect data. For example, this can be executed primarily through a questionnaire at the initial contact with the participant. There are several questionnaires available in the literature such as the Motion Sickness Susceptibility Questionnaire (MSSQ) [165, 166] and the Virtual Reality

[17] https://looxidlabs.com/
[18] https://www.emotiv.com/epoc/
[19] https://neurable.com/
[20] https://openbci.com/
[21] https://choosemuse.com/

Sickness Questionnaire (VRSQ) [167]. The reader may refer to the literature (MSSQ [82] and VRSQ [83]) for in detail usage.

As the second, having participants prior experience with an immersive VR environment is of great importance. For example, some research findings may be influenced by the novelty bias caused by first-time excitement in the VR medium [18, 20]. This can be further explained by the discrepancies between expected and reported emotions. Furthermore, due to the highly immersed VR content, participants tend to report a positive category, even though the material is expected to trigger a negative experience. On another note, it would be more difficult to find emotional content that induces negative emotions. The impact of the first-time VR experience can be minimized by employing participants with prior VR experience or by conducting a training session [105] with targeted participants prior to data collection. This provides additional advantages as, participants are aware of the perspective of a VR environment, and have an overall understanding of the hardware mechanics, especially while playing games using VR controllers. This further reduces the stimulation of unexpected emotions such as frustration when the exact mechanism is not known. Also, by employing this practice, researchers can recruit participants who are less vulnerable to motion sickness.

Another constraint while employing VR in affective research is the limited ability to study full facial expressions. Although there are a few growing interfaces for measuring facial expressions via physiological signals, their potential is limited to a few signal points and hence do not account for measuring detailed facial features. Most importantly, to conduct credible research, the applicability of using those hardware interfaces incorporated with diverse facial structures (i.e., fitting of interfaces) of participants should be verified for effective future directions.

## 10 Discussion

The definition of emotions has been obtained from various studies through multidimensional data collection settings. While this understanding of definitions is valid for historical understanding, many studies do not fully account for the natural and realistic emotion generation. As such, most emotional studies focus on passive emotional elicitation, and the participant is an observer of the overall emotional encounter. In that aspect, contemporary studies have directed and focused more on the formation of realistic virtual settings that increase the efficacy of evoked emotion, enhancing engagement and improving immersion by VR technology.

So, in this endeavour, a survey has been conducted on the potential of using virtual reality as a medium to elicit discrete and dimensional models of emotions. Here we mainly discussed visual and audio-visual, games, tasks, 360-degree panoramas and 360-degree videos as main media content. We systematically reviewed public datasets generated using VR for emotional studies, familiarizing researchers with literature relevant to their domain, and directing researchers for the induction of realistic emotions by VR. In this review, we also analysed the contemporary emotion-sensing interfaces which can be used for more accurate acquisition of physiological signals and facial expression in affective computing studies designed by a VR protocol. Therefore, it is expected that these commercialized wearable devices will disclose novel research directions by mitigating the inability of facial expression analysis due to the occlusion of the facial activities by the HMD. More, we presented a review of subjective measurement approaches and limitations of using VR in emotional research.

The followings are the basic conclusion of our study on VR to evoke emotions:

1. Using VR to study emotions, psychological traits and HCI is an elevated trend when formulating experimental procedures.
2. VR based media content are accountable for the elicitation of emotions and study both discrete and dimensional models.
3. The state-of-the-art research has excelled in using visual and audio stimuli, games, 360-degree material in combination with VR and has been able to elicit a wide range of emotions.
4. Even though VR games-directed research is limited, it can be used to study different domains compared to other major media types due to the first-person role-play, virtual collaboration, and being an empathy machine.
5. VR provides the possibility to trigger changes in physiology, consequently, researchers can effectively assess bodily functions and reactions to investigate the emotional-physiology relationship.
6. VR games can trigger motivational tendencies to win. These features of games bring the ability to effectively access the motivational aspects of an experience. So, games can be used to evaluate emotional models like Component Process Model (CPM) [53] which treat Motivation as a major component and explore motivation tendencies for an extended time [117].
7. VR games can be used to evaluate rapidly changing physiological models that are defined by assuming a temporal perspective on emotion generation.

Considering the importance of emotions to humanity and society and their particular role in creating a natural human-computer interaction, emotion researchers should consider the following for the proliferation of a reliable model for effectively understanding emotions:

1. Many published examinations have assumed a passive elicitation of emotions and were unable to obtain the active involvement of participants in the data collection process. However, an increasing number of studies have empirically shown that active emotion elicitation methods are more pragmatic in evoking natural feelings. Therefore, a novel direction must be initiated with an active data collection mechanism such as VR to acquire a more reliable explanation in Affective Computing.
2. Benchmark datasets are required to study emotions and make consistent comparisons across different studies. In the long run, this will pave the path for the development of more generalized models that can arrive at a consensus on the evolving debates about the conceptualization of basic emotions and the mechanisms involved in emotion formation.
3. There are relatively few publicly accessible databases based on VR in the literature. As we explained earlier, VR involves the active collaboration of participants in the experiments which is a greater advantage in triggering real emotions. However, the limitation of publicly accessible datasets obstructs the development of novel research directions due to the lack of resources. Therefore, we noted the importance of establishing public datasets.

4. Existing VR contents are more suitable for eliciting positive emotions and there are limited VR contents to study negative emotions.
5. Future research can also consider the usage of emotion-sensing interfaces (Section 8) with VR headsets to measure facial expressions and physiological signals.
6. Few recent research has focused on eliciting multimodal emotions through haptic feedback [104, 168], which is a signpost to a novel direction. However, the domain is still immature, and the data collection methodology requires in-depth investigations to generate more intense emotion.
7. In AC, very limited research has adopted the principles of Appraisal theory [52] and the Component Process Model [53], even though these models can better reconcile with the recent neuroscience findings on emotion [51]. Therefore, we identified the importance of considering such models in future research. Other important theories of emotion such as the theory of constructed emotion [169] has also been neglected by the AC community so far and needs to be incorporated into the current approaches.

## 11 Conclusion

To our knowledge, this study is the very first survey of Virtual Reality to induce emotions. We conducted this survey by analysing the literature and we showed the possibility of novel VR technologies to elicit emotions for affective computing studies and potentially beyond. We also discussed the possibility of VR to elicit discrete and dimensional models of emotions using visual and audio-visual stimuli, games, and 360-degree panoramas and videos as emotive content. Public datasets generated via VR and sensors that can be used in future research were presented. Although it is ahead of the scope of this work, the conclusions imply potential future real-time adaptive VR design based on users' emotions. With the continuous evolution of the experimental design with various data elicitation mechanisms, the understanding and classification of emotion will lead and enhance functions with computerized systems based on Affective Computing competencies.

# Supplementary Material

## Supplementary material A

Table 3 An assessment of visual and audio-visual VR material usage in Affective Computing, psychology, and Human-Computer Interaction. (PPG: Photoplethysmogram, EDA: Electrodermal Activity, EEG: Electroencephalography, VE: Virtual Environment, SVM: Support Vector Machine, KNN: K-Nearest Neighbour, NB: Naïve Bayes, RF: Random Forest, EMG: Electromyography)

| Reference | Focus | Purpose of using VR | N participants | Age | Material | VR device | Method |
|---|---|---|---|---|---|---|---|
| Li, et al. [91] | Investigate the relation of head yaw with valence and arousal ratings | Emotion elicitation | 95 | 18-24 | Videos | Oculus Rift | Pearson's product-moment correlations |
| Mavridou, et al. [97] | Arousal detection via PPG | Emotion elicitation | 11 | 18-35 | Videos | Faceteq | Machine Learning (SVM) |
| Felnhofer, et al. [106] | Elicit joy, sadness, boredom, anger, and anxiety emotions, Analyse presence over several VEs, Relation of arousal, EDA, and Presence | Emotion elicitation | 120 | Mean of 24.89 | VE | Sony HMZ-T1 3D Visor | Statistical Analysis, Pearson Correlations |
| Bilgin, et al. [128] | Analyse VR and 2D display across EEG | Emotion elicitation and regulation | 10 | 21-35 | VE | OSVR HDK2 VR | Machine Learning (SVM) |
| Cebeci, et al. [82] | Investigate effects of VE on different contexts | Cybersickness elicitation | 20 | Mean of 27 pm 8.46 | VE | HTC Vive | Statistical Analysis |
| Nam, et al. [90] | Increase engagement in VR with biosignals | Emotion elicitation | 13 | 22-34 | Images | HTC Vive | Machine Learning (SVM) |
| Dey, et al. [104] | Impact of physiological feedback on emotions evoked in VE | Triggering physiological feedback in VR | 19 | 21-45 | VE | HTC Vive | Statistical Analysis |
| Mavridou, et al. [116] | Introduce EmteqVR wearable technology | Emotion elicitation | 400 | Not given | Scenes | EmteqVR | N/A |
| Riva, et al. [25] | Analyse the possible use of VR as an affective medium and the relationship between presence and emotions | Emotion and presence elicitation | 61 | 19-25 | VE | Not given | Statistical Analysis |
| Zhang, et al. [67] | Design an affective VR system by scenes annotated by valence, arousal, and dominance | Emotion elicitation | 100 | Not given | Scenes | Not given | N/A |
| Liao, et al. [95] | Design an emotion elicitation system and evaluate arousal | Emotion elicitation | 31 | 18-22 | Scenes | HTC Vive | Statistical Analysis |
| Hinkle, et al. [56] | Investigate the subjects' emotional response to stimuli, the possibility to measure physiological signals in VR environments and discuss the challenges. Develop a standard feature extraction criteria | Emotion elicitation | 5 | 20-50 | Videos and Games | Oculus Rift | Statistical Analysis, Machine Learning (NB, KNN, SVM) |
| Estupiñán, et al. [22] | Investigate the possibility of VR to improve the | Emotion elicitation | 10 | 23-31 | Images | Oculus Rift | Statistical Analysis |
| | emotional experience by images | | | | | | |
| Gutiérrez-Maldonado, et al. [21] | Generate dynamic VR faces that can integrate into VR cyber interventions | Create avatars and social environments | 98 | 18-65 | Avatar | Acer Aspire computer, 3D glasses | Statistical Analysis |
| Bekele, et al. [57] | Generate and evaluate a VR based system for emotion analysis in facial expressions for patients with schizophrenia (SZ) | Emotion elicitation | 12 (healthy) and 12 (SZ) | Mean healthy 44.9, mean SZ 45.7 | Avatar | 24 flat LCD panel monitor | Statistical Analysis, Machine Learning (Clustering) |
| Malbos, et al. [170] | Investigate the effectiveness of VR to treat Squalophobia | Triggering Squalophobia | 1 | Not given | VE | Sony PSVR | Statistical Analysis |
| Marín-Morales, et al. [171] | Investigate the physiological activities while participants exploring VEs | Emotion elicitation | 45 | Mean of 28.9 | VE | HTC Vive | Statistical Analysis |
| Kim, et al. [26] | Investigate the usage of VR as an affective medium, and the correlation of immersion and emotion | Emotion elicitation | 38 | Not given | Videos | HTC Vive | Statistical Analysis |
| Chirico, et al. [110] | Evaluate the potential of VR in inducing awe and compare the efficacy of VR and 360º-videos | Awe elicitation | 36 | Mean of 23.33 | VE | Oculus Rift | Statistical Analysis |
| Myung and Jun [172] | Analyse physiological responses by evoking plan configurations through VR | Emotion elicitation in housing plan configurations | 60 | Mean of 22 | VE | Oculus Rift | Statistical Analysis |
| Gnacek, et al. [173] | To study the possibility of using PPG sensors integrated into VR headset to measure the heart rate | For evaluating the reliability of the PPG sensor system embedded in a headset | 16 | Mean of 25.81 | VE | HTC Vive | Statistical Analysis |
| Xu, et al. [68] | To develop a framework for EEG based emotional study by VR scenes | Emotion elicitation | 19 | 19-25 | Scenes | HTC Vive | Machine Learning (GBDT, RF, SVM) |
| Zhao, et al. [123] | To propose a live storytelling emotion embodiment framework with facial expressions and auxiliary multi-modalities | Emotion embodiment | 6 | 16-22 | Avatar | Oculus Rift | Quantitative analysis |
| Schmidt, et al. [174] | To formulate a human-agent interaction framework | Agent interaction | 40 | 18-41 | Avatar | CAVE | Qualitative analysis |
| Thejaswini S [175] | To propose a hybrid LSTM framework by channel fusion of EEG data in differentiating discrete and dimensional emotions | Emotion elicitation | 66 | Mean of 35 | Videos | Not given | Deep Learning (LSTM) |
| Dey, et al. [120] | To understand presence in VR using physiological and neurological signals | Evoking presence | 24 | 20-30 | VE | HTC Vive | Statistical Analysis |
| Simo Järvelä [70] | To investigate the impact of neurofeedback in shared VR space. | Sharing neurofeedback in social interactions | 78 | 19-50 | VE | Oculus Rift | Statistical Analysis |
| Li, et al. [122] | To design an affective VR environment with biofeedback. | Emotion elicitation | 6 | 9-14 | VE | Oculus Rift | Statistical Analysis |
| Mavridou, et al. [100] | Valence detection from facial EMG | Emotion elicitation | 34 | 18-40 | Visual (videos) | Faceteq | Machine Learning (SVM) |
| Hofmann, et al. [96] | Arousal detection from EEG | Emotion elicitation | 45 | 20-32 | VE | HTC Vive | Deep Learning (LSTM) |

## Supplementary material B

Table 4 An assessment of games and tasks VR material usage in Affective Computing, psychology, and Human-Computer Interaction. (SVM: Support Vector Machine, KNN: K-Nearest Neighbour, DA: Discriminant Analysis, RF: Random Forest, GBoT: Gradient Boosting of trees, GPR: Gaussian Process Regression, GLM: Generalized Linear Model, GB: Gradient Boosting, NB: Naïve Bayes, DT, Decision Trees)

| Reference | Focus | Purpose of using VR | N participants | Age | Material | VR device | Method |
|---|---|---|---|---|---|---|---|
| Granato, et al. [19] | Investigate the player emotions by physiological signals | Emotion elicitation | 33 | 18-40 | Games | Oculus Rift | Machine Learning (SVM, RF, GBoT, GPR) |
| Susindar, et al. [69] | Investigate the impact of presentation medium on decision making | Decision making in an emotional situation | 11 | 19-29 | Games and Tasks | Oculus GO | Statistical Analysis |
| Peng, et al. [20] | Investigate the impact of the emotional challenge on player experience with and without conventional challenge in VR and PC | Emotional challenge induction | Study 1:28 Study 2:42 | Study 1: mean 24.4 Study 2: mean 23.7 | Games | HTC Vive | Statistical Analysis |
| Meuleman and Rudrauf [18] | Investigate the potential of VR to evoke intense emotions and to study the Component Process Model of emotions | Emotion elicitation | 53 | Mean of 28.7 | Games | HTC Vive | Machine Learning (Supervised, Unsupervised) |
| Hart, et al. [176] | Understand the possibility to capture and share facial expressions between players | For sharing facial expressions in a cooperative environment | 4 | Mean of 27 | Games | Samsung Odyssey | Statistical Analysis |
| Bassano, et al. [107] | Elicit joy and frustration emotions assuming the appraisal model of emotions | Emotion elicitation | 5 | Not given | Games and Tasks | HTC Vive | N/A |
| Lin [80] | Investigate coping strategies and fright responses in VR horror games | Fear elicitation | 145 | Mean of 22.57 | Games | HTC Vive | Statistical Analysis |
| Pallavicini and Pepe [75] | Investigate the potential of VR games to induce positive and reduce negative emotions and to identify the impact of body involvement | Emotion elicitation and body involvement | 36 | Mean of 25.6 | Games | HTC Vive | Statistical Analysis |
| Reidy, et al. [85] | Introduce a cognitive training system via VR gamification tasks | Cognitive training | 18 | 20-37 | Tasks | Faceteq | Machine Learning (SVM, KNN, LDA) |
| Caldas, et al. [83] | Investigate the impact of presence and challenge on emotional engagement in VE | Triggering presence and challenge in | 87 | Mean of 24.64 | Games | Oculus Rift | Statistical Analysis |

| | | | | | | | | |
|---|---|---|---|---|---|---|---|---|
| | | an emotional state | | | | | | |
| Moghimi, et al. [105] | Design an affective computing system | Emotion elicitation | 30 | Mean of 22.76 | Games | Samsung HD LCD display | Machine Learning (KNN, SVM, DA, Tree) |
| Gupta, et al. [81] | Investigate the possibility of VR to induce and evaluate trust under both high and low cognitive load | Triggering trust in cognitive load situations | 24 | 23-35 | Tasks | HTC Vive | Statistical Analysis |
| Bălan, et al. [58] | Define a methodology to recognize artefacts in VR gameplay based on phobia | Evoking acrophobia | 5 | 24-50 | Games | HTC Vive | Statistical Analysis |
| Smeijers and Koole [117] | Investigate the impact of VR games on the manipulation of aggression and explore the effects of motivational intervention | Manipulation of aggression | Not given | Not given | Games | Not given | Statistical Analysis |
| Li, et al. [78] | Introduce an affective design with VR and biosignals measures | Emotion elicitation | Not given | Not given | Games | Oculus Rift | Graph-based modeling |
| Lipp, et al. [111] | Investigate the emotional experience in freeze-frame objects via VR simulator | Emotion elicitation in freeze-frame | 60 | 19-40 | Tasks | HTC Vive | Statistical Analysis |
| Kim, et al. [177] | To understand the correlation of body positions and sense of presence in VR | Generate different body positions in several freedom levels | 62 | Not given | Games | Oculus Rift | Statistical Analysis |
| Song, et al. [178] | To investigate the efficacy of VR and AR games in easing neck pain by mobile phone usage | For alleviating neck pain | 24 | 19-25 | Games | Not given | Statistical Analysis |
| Steinhaeusser and Lugrin [112] | To compare the impact of medium (SSR, VR, PC) on horror games and investigate the sense of presence and emotional intensity | Horror elicitation | VR-20 | Mean of 25.10 | Games | HTC Vive | Statistical Analysis |
| Kano and Morita [121] | To study the factors that generate empathetic behaviour in virtual agents | Embodiment in empathy situations | 32 | Not given | Games | HTC Vive | Statistical Analysis |
| Giglioli, et al. [179] | To analyse the applicability of virtual serious games to study personality needs. | Personality recognition | 61 | 18-55 | Games | HTC Vive | Machine Learning (Conditional inference trees, RF, SVM, GLM) |
| Delvigne, et al. [139] | To propose a framework that records biosignals and predict the attention state. | Attention analysis | 32 | 19-30 | Tasks | HTC Vive Pro | Machine Learning/ Deep Learning |
| Pallavicini, et al. [109] | To compare the player experience in VR and desktop game sessions. | For player experience analysis | 24 | 18-35 | Games | Oculus Gear | Statistical Analysis |
| Oberdörfer, et al. [124] | To compare the impact of immersion on decision making. | For immersion in decision making | 25 | Mean of 20.44 | Tasks | HTC Vive Pro | Statistical Analysis |
| de-Juan-Ripoll, et al. [119] | To develop a framework that can differentiate participants by personality, sensation seeking, behaviour and | Risk environments elicitation | 98 | Mean of 37.08 | Games | HTC Vive Pro eye | Statistical Analysis/ Machine Learning |
| | biosignals in risk environments. | | | | | | |
| Wilson and McGill [108] | To compare the impact of VR and non-VR games and investigate the applicability of prevailing game ratings. | For player experience analysis | 16 | 18-30 | Games | PlayStation | Statistical Analysis |
| Amores, et al. [164] | Arousal detection from EEG | Emotion elicitation | 18 | 23-49 | Scenes and games | HTC Vive Pro | Deep Learning (CNN) |
| Shumailov and Gunes [102] | Affect recognition in VR using EMG signals collected from arms. | Emotion elicitation | 8 | Not given | Games | HTC Vive | Machine Learning (SVM) |
| Ishaque, et al. [113] | To investigate stress by VR games | For alleviating stress | 14 | 20-40 | Games | Not given | Machine Learning (DA, DT, SVM, GB, NB) |

## Supplementary material C

Table 5 An assessment of 360º panoramas and videos VR material usage in Affective Computing, psychology, and Human-Computer Interaction. (EEG: Electroencephalography, SVM: Support Vector Machine, KNN: K-Nearest Neighbour, NB: Naïve Bayes, RF: Random Forest, GBM: Gradient Boosting Machine, DNN: Deep Neural Network)

| Reference | Focus | Purpose of using VR | N participants | Age | Material | VR device | Method |
|---|---|---|---|---|---|---|---|
| Marín-Morales, et al. [92] | Investigate emotional state elicited via VR by ML techniques | Emotion elicitation | 38 | Mean of 28.42 | 360º panorama | Samsung Gear | Machine Learning (SVM) |
| Teo, et al. [93] | Implement a mixed reality collaboration mechanism to share the 360º panorama with a remote user | Remote collaboration | 22 | 18-40 | 360º panorama | HTC Vive | Statistical Analysis |
| Teo, et al. [94] | Introduce a mixed reality collaboration approach with 360º panorama and 3D constructed scenes | Remote collaboration | 20 | 18-45 | 360º panorama | Samsung Odyssey | Statistical Analysis |
| Suhaimi, et al. [77] | Establish a 360º VR video dataset for emotional analysis | Emotion elicitation | 15 | 21-41 | 360º video | Not given | Statistical Analysis |
| Gupta, et al. [101] | To develop a personalised emotion recognition system | Emotion elicitation | 6 | Mean of 34.83 | 360º video | HTC Vive | Machine Learning (NB, KNN, SVM, RF) |
| Toet, et al. [103] | Validate a graphical affective reporting scale | Emotion elicitation | 40 | 18-29 | 360º video | Samsung Odyssey | Statistical Analysis |
| Nazmi Sofian Suhaimi [114] | Four class emotion classification via EEG data | Emotion elicitation | 31 | Not given | 360º video | Not given | Machine Learning (KNN, SVM) and Deep Learning |
| Aaron Frederick Bulagang [133] | Investigate the feasibility of using ECG to predict emotions | Emotion elicitation | 5 | 20-28 | 360º video | HTC Vive | Machine Learning (SVM) |
| Vallade, et al. [136] | To study the possibility of increasing public speaking via 360º videos | Training public speaking | 86 | 18-29 | 360º video | Oculus Go, Oculus Quest | Quantitative analysis |
| Zheng, et al. [134] | Predict four classes of emotions via pupillometry | Emotion elicitation | 10 | 21-28 | 360º video | HTC Vive | Machine Learning (SVM, kNN, RF) |
| Bernal, et al. [154] | Introduce PhysioHMD as an interface that can be integrated with VR, AR headset to collect biometric data | Emotion elicitation and analysis wearable device for VR headset | 8 | 18-32 | 360º scene and video | HTC Vive | Deep Learning (CNN) |
| Suhaimi, et al. [115] | Emotion classification using EEG | Emotion elicitation and analysis of wearable EEG | 31 | 23-28 | 360º video | VR Box | Machine Learning (RF, GBM, NB) |
| Teo and Chia [99] | Investigate the possibility of wearable EEG in emotion recognition. | Emotion elicitation | 24 | 20-28 | 360º video | Not given | Deep Learning (DNN) |

## Supplementary material D

Table 6 An overview of VR games used in literature with emotional experience

| Reference | Game | Description | Dominant emotions experienced |
|---|---|---|---|
| Granato, et al. [19] | Project Cars[22] | A racing game on existing car models | Valence, arousal |
| Granato, et al. [19] | RedOut[23] | A racing game on futuristic shuttle models | Valence, arousal |
| Susindar, et al. [69] | Play With Me[24] | An adventure in a mysterious house | Fear |
| Peng, et al. [20] | War Never Changes from Fallout 4[25] | A story of a family facing a sudden nuclear war | Anxiety, sadness, helplessness, worry, shock, powerlessness, trust, relaxation, surprise, guilt, tension, despair, love, empathy |
| Peng, et al. [20] | When Freedom Calls from Fallout 4 | A sudden entering of a player to a building with gunmen and a fight afterwards | Tension, amusement, excitement, courage, fear, stress, empathy |
| Peng, et al. [20] | Out of the Fire from Fallout 4 | The game starts with a conversation between father and son. Then son fight against enemies to take back heirloom sword | Hope, courage, love, tension, worry, anxiety, stress, relief, amusement, trust, contentment, excitement, joy, empathy, delight, pleasure |
| Peng, et al. [20] | Out of the Fire from Fallout 4 without conversations, strong characters | A story without conversations. Player needs to fight against enemies | Amusement, delight, relaxation, interest, fear, excitement, joy contentment |
| Meuleman and Rudrauf [18] | Tilt Brush[26] | A painting experience in a virtual palette | Interest, amusement, pleasure, joy, contentment |
| Meuleman and Rudrauf [18] | theBlu - Whale[27] | An underwater sea life experience | Interest, admiration |
| Meuleman and Rudrauf [18] | Fruit Ninja VR[28] | A game to slash fruits with swords | Amusement, pleasure, interest, joy |

[22] https://www.projectcarsgame.com/
[23] https://34bigthings.com/portfolio/redout/
[24] https://www.oculus.com/experiences/go/1226116427458674/
[25] https://fallout.bethesda.net/en/games/fallout-4
[26] https://store.steampowered.com/app/327140/Tilt_Brush/
[27] https://store.steampowered.com/app/451520/theBlu/
[28] https://store.steampowered.com/app/486780/Fruit_Ninja_VR/

| | | | |
|---|---|---|---|
| Meuleman and Rudrauf [18] | The Lab – Longbow[29] | A shooting game against enemies using a bow | Pleasure, amusement, joy |
| Meuleman and Rudrauf [18], Lin [80] | The Brookhaven Experiment[30] | A shooting game against zombies | Fear |
| Meuleman and Rudrauf [18] | Richie's Plank Experience[31] | An experience from a taller building | Fear |
| Meuleman and Rudrauf [18] | Zero G[32] | A zero-gravitation experience | Fear, interest, contentment |
| Hart, et al. [180] | Bomb Defusal (developed by authors) | A bomb defusal experience in collaboration with a remote player | Fearful and monster facial expressions rendered by an avatar |
| Hart, et al. [180] | Island Survivor (developed by authors) | An escape from a remote island in collaboration with a remote player | Fearful and monster facial expressions rendered by an avatar |
| Bassano, et al. [107] | Kitty Rescue[33] | An experience to save kittens from danger | Fear |
| Bassano, et al. [107] | Suit assembling (developed by authors) | A time manipulated task to assemble an exoskeleton | Frustration |
| Bassano, et al. [107] | Shinrin-yoku: Forest Meditation and Relaxation[34] | An interactive forest exploration with trees, animals | Joy, awe, relief |
| Bassano, et al. [107] | RideOp - VR Thrill Ride Experience[35] | A thrill experience by riding in world's attractions places | Fear, awe |
| Pallavicini and Pepe [75] | Fruit Ninja VR | A game to slash fruits with swords | Happiness, surprise |
| Wilson and McGill [108] | Resident Evil 7 | A survival horror game with bad language, violence, and attacks | Fear |
| Hinkle, et al. [56] | InCell[36] | A game inside human cells to stop virus propagation | Excitement, pleasant |
| Hinkle, et al. [56] | Lucky's Tale[37] | A third-person adventure game with puzzles and thrilling challenges | Excitement, pleasant |
| Shumailov and Gunes [102] | Egg Time[38] | A game to catch eggs and defuse bombs | Valence, arousal |
| Shumailov and Gunes [102] | Google Earth VR[39] | An interactive exploration of the Earth | Valence, arousal |
| Shumailov and Gunes [102] | Spell Fighter VR[40] without speech input | A sword fight against skeletons where the player needs to cast spells. But authors have used the game without speech input. | Valence, arousal |
| Shumailov and Gunes [102] | The VR Museum of Fine Art[41] | An exploration of a museum | Valence, arousal |
| Shumailov and Gunes [102] | Fruit Ninja VR | A game to slash fruits with swords | Valence, arousal |
| Pallavicini, et al. [109] | Smash Hit[42] | A shooter game to hit obstacles and collect spheres | Happiness, surprise |

[29] https://store.steampowered.com/app/450390/The_Lab/
[30] https://store.steampowered.com/app/440630/The_Brookhaven_Experiment/
[31] https://store.steampowered.com/app/517160/Richies_Plank_Experience/
[32] https://store.steampowered.com/app/567890/ZeroG/
[33] https://store.steampowered.com/app/726620/Kitty_Rescue/
[34] https://store.steampowered.com/app/774421/Shinrinyoku_Forest_Meditation_and_Relaxation/
[35] https://store.steampowered.com/app/972890/RideOp__VR_Thrill_Ride_Experience/
[36] https://www.oculus.com/experiences/rift/814255758700053/
[37] https://www.oculus.com/experiences/rift/909129545868758/
[38] https://store.steampowered.com/app/531990/Egg_Time/
[39] https://store.steampowered.com/app/348250/Google_Earth_VR/
[40] https://store.steampowered.com/app/455440/Spell_Fighter_VR/
[41] https://store.steampowered.com/app/515020/The_VR_Museum_of_Fine_Art/
[42] https://www.oculus.com/experiences/gear-vr/942006482530009/

## Supplementary material E

Table 7 A list of studies that used VR to investigate discrete and dimensional models of emotions, their materials, content types, outcomes, and collected annotations or measures. (VE: Virtual Environment, HMD: Head Mounted Display, SSR: Smart Substitutional Reality, EDA: Electrodermal Activity, EMG: Electromyography, ACC: Acceleration, BVP: Blood Volume Pulse, GSR: Galvanic Skin Response, IBI: Inter-Beat-Interval, TEMP: Temperature, HR: Heart Rate, EEG: Electroencephalography, SAM: Self-Assessment Manikin, EOG: Electrooculography, ECG: Electrocardiography, kNN: K-Nearest Neighbor, NB: Naïve Bayes, SVM: Support Vector Machine, SC: Skin Conductance, AUC: Area Under the Curve, PPG: Photoplethysmogram, DNN: Deep Neural Network)

| Reference | Material | Content/story | Outcome | Annotations/Measures |
|---|---|---|---|---|
| Liao, et al. [95] | Scenes were generated from images and sound. | Peace: Rainforest scenery and villa tour<br>Sad: Earthquake<br>Happy: Fairy tale castle<br>Fear: An empty cave, horror palace, abandoned house<br>Distaste: Chaotic house | Heart rate analysis showed that the mean arousal of VR was lower than that of video, and therefore VR and video were not significant from the point of view of arousal.<br><br>SAM scorings illustrate a significant difference between VR and video only to fear. | Pleasure, happy, fear distaste<br><br>Arousal<br><br>HR |
| Xu, et al. [68] | Scenes | Happy, fear, peace, disgust VR scenes | Accuracy: 81.3%. | Valence, arousal, dominance<br><br>EEG |
| Felnhofer, et al. [106] | VEs | Joy: Pleasant, calm, quiet, sunny background with birds<br>Sadness: Gray and rainy day with non-players walking quickly in pathways<br>Anger: A construction site VE having constant noise of hatred and frustration<br>Anxiety: Gloomy, dark night scenario<br>Boredom: An empty scenario without non-players, empty benches, and trees | Ability to elicit intended emotion by each of five scenarios.<br>The level of presence was similar across all VEs. | Joy, sadness, anger, anxiety, boredom<br><br>Presence<br><br>EDA |
| Riva, et al. [25] | VEs | Relax: A nature park<br>Anxiety: A park in a dark background<br>Neutral: | The efficacy of using VR to induce anxiety and relaxation is proved.<br>Observed a higher level of presence in anxiety VE. | Happiness, sadness, anger, surprise, disgust, anxiety, quietness<br><br>Level of anxiety |
| Chirico, et al. [110] | VEs | Forest and waterfall environment, snowy mountain and wind content, and an Earth view as an astronaut | All three environments indicated significant differences compared to neutral and were differed from the baseline.<br>A higher level of awe was observed in mountain scenarios related to forests and the Earth. | Anger, disgust, fear, pride, amusement, sadness, joy, awe |
| Malbos, et al. [170] | VEs | Virtual swimming pool with sharks | Revealed self-reported fear and anxiety | Mood, depression, anxiety state, quality of life, Squalophobia level |
| Dey, et al. [104] | VEs | A story with dangerous animals as panthers, dinosaurs and attacking snakes | Discovered self-reported interest, excitement, scared, nervousness, afraid, and providing manipulated heart rate can change the emotional state. | Interest, excited, scared, nervous, afraid.<br><br>Valence, arousal, dominance |
| Cebeci, et al. [82] | VEs | A horrific hospital, that has run-down surroundings, terrifying creatures, and emotionally unpleasant scenarios with audio, a neutral campfire environment, a roller-coaster environment that can expect cybersickness. | Cybersickness causes a higher saccade speed.<br><br>Observed a decrease in self-reported comfort and happiness after the experience of unpleasant and cybersickness scenarios | Happy, sad, comfortable, anxious<br><br>HR, eye tracking |
| Lipp, et al. [111] | VEs | Car accident with freeze-frames in the middle of the simulation | Cohen's rating of third study, fear: 0.81, anger: 0.62, guilt:0.78, sadness: 0.66. | Joy, love, fear, anger, guilt, sadness |
| Simo Järvelä [70] | VEs | Nature-based shared social environment followed by an evening meditation session in a shrine | Observed higher intensities of valence and excitation in the dyadic setting, while higher valence was observed in the solitary state. | Empathy, sympathy, compassion, soft-hearted, warm, tender, moved.<br><br>Valence, arousal, dominance<br><br>Respiration, EEG |
| Marín-Morales, et al. [171] | VEs | Real museum and a virtual museum in low and high arousal conditions | Reported physiological changes in the real scenario but not in the virtual scenario for high and low arousal conditions. | ECG |
| Hofmann, et al. [96] | VEs | Rollercoasters | Arousal classification accuracy 75.7% | EEG, SC, HR<br><br>Arousal |
| Kim, et al. [26] | Videos | A film expert from the Korean movie "Gidam" for horror, a scene of Ellie and Carl's relationship from the movie "Up" for empathy | Revealed more intense fear in VR than traditional experiences. (Higher level of emotional experience with a mean of 5.22 (SD = 0.92) and a mean of 4.29 (SD = 0.79) for HMD and non-HMD conditions respectively)<br><br>Observed self-reported sadness, but no significant between HMD and non-HMD environments when watching empathy movies | Fear, anxiety, disgust, surprise, tension, happiness, sadness, love, being touched, compassion, distressing, disappointment.<br><br>Immersion |
| Mavridou, et al. [97] | Videos | Valence and arousal videos triggering the 4 quadrants of the dimensional model, neutral videos | AUC using PPG and EEG signals = 0.69 | Arousal |
| Mavridou, et al. [100] | Videos | Valence and arousal videos triggering the 4 quadrants of the dimensional model, neutral videos | Valence detection rate from the facial EMG using SVM: 82.5% | Valence |
| Ding, et al. [98] | Scenes and games | Low arousal - Frozen Lake with snow and birds including music.<br>High arousal – An adaptive stone avoidance game where the speed of stone movement is controlled by the subject's scores. | Accuracy of arousal detection from CNN: 86.03% | Arousal |
| Hinkle, et al. [56] | Games and videos | The Rose and I movie, Two discovery action videos, | SVM performs better than kNN and NB with an | Valence, arousal |
| | | InCell game, Lost movie, Oculus Dream Deck, Luckys Tale game | accuracy of 80% for domain-specific and 89.19% for proposed general features of physiological signals. | EEG, EOG, EMG, EDA, ECG, Respiration, TEMP, HR, BVP, Blood oxygen, head acceleration and rotation, body acceleration and rotation |
| Bassano, et al. [107] | Games and tasks | Amuse and awe content, meditation, and relaxing experiences, a time manipulated suit assembling task, where the counting timer finally drops, and the player is impossible to complete and a thrill ride scenario | This preliminary study showed self-reported positive and negative emotions. | 16 discrete emotions<br><br>Speech, EMG, ACC, BVP, GSR, IBI, TEMP |
| Pallavicini, et al. [109] | Games | First-person shooter game to hit obstacles and collect spheres (Smash Hit game) | Revealed the experience of higher levels of happiness, surprise, HR, skin conductance in VR games compared to desktop computers.<br>Reported similar performance based on the total time in comparison to VR and desktop. | Anxiety, happiness, surprise |
| Peng, et al. [20] | Games | A game having an emotional challenge based on a sudden nuclear war faced by the family ("War Never Changes" in Fallout 4 game), a first-person shooting against enemies, ("When Freedom Calls" in Fallout 4 game), a gunfight with emotional conversations regarding the family background, a gunfight without emotional conversations | Emotional challenge triggered basic negative emotions and when included more emotional challenge it induced complex positive and negative emotions.<br>Observed significantly higher presence in VR condition than the desktop. | 48 discrete emotions<br><br>challenge, usability, appreciation, enjoyment, suspense, immersion, presence |
| Moghimi, et al. [105] | Games | Time-limited and performance-based speed boat simulation | F1-score of EEG, GSR, and HR signals in identifying relax: 95.50%, content: 95.06%, happy: 94.12%, excited: 95.61%, angry: 94.4%, afraid: 93.11%, sad: N/A, bored: 94.87% | Valence, arousal dominance<br><br>Relaxed, content, happy, excited, angry, afraid, sad, bored |
| Pallavicini and Pepe [75] | Games | Slice fruits by swords (Fruit Ninja game), a dance game in which player needs to hit the orbs (Audioshield game) | Statistically significant increase in happiness and decrease in negative emotions as fear and sadness. | Happiness, surprise fear, sadness, anxiety |
| Lin [80] | Games | A dark survival scenario from zombies where the player has a gun with limited bullets (Brookhaven Experiment game). | Revealed a significantly higher fear experience from self-reports. | Fear |
| Meuleman and Rudrauf [18] | Games | Virtual painting game (Title Brush game), Sealife experience (The Blu game), a game to slice fruits by swords (Fruit Ninja game), shooting game against enemies (The Lab), a shooting game against zombies (The Brookhaven Experiment game), a virtual-navigation in a taller skyscraper (Richie's Plank | Revealed self-reported 20 discrete emotions and identified a cluster of fear and joy in the componential space of emotions. | 20 discrete emotions |
| | | Experience game), a virtual-navigation in zero-gravity (Zero G game) | | |
| Steinhaeusser and Lugrin [112] | Games | The first-person role of a journalist examining mysterious events in an abandoned laboratory. The player must communicate with the ghosts to solve puzzles and fight off the monsters. | Did not find any difference in the emotional intensity of the three mediums (VR, PC, SSR), the results showed that VR and SSR were at a higher level of presence compared to the PC. | Anxiety level, Fear |
| Bălan, et al. [58] | Games | An acrophobia (fear of heights) scenario in a mountain landscape | A system to treat acrophobia patients were proposed. | Fear<br><br>EDA, HR, Respiration |
| Wilson and McGill [108] | Games | First-person survival violent game "Resident Evil 7" with harsh language. | Higher presence and embodiment in violent VR game sessions in comparison to non-VR. | Anxiety level, Presence |
| Susindar, et al. [69] | Games and videos | A decision-making scenario under the impact of emotions ("Play with me" game) | Reported that VR has the efficacy to study fear and anger while decision making | Fear, anger |
| Reidy, et al. [85] | Games | Collecting products from a supermarket (easy, medium, hard levels) and multi-room museum | Accuracy of valence: 64.1%, arousal: 76.2% | Valence, arousal<br><br>f-EMG |
| Caldas, et al. [83] | Games | A first-person role-playing skydiving game | Reported that presence rich content is related to arousal and challenge is associated with valence and dominance. Respiration and SC reported increments during high challenged content and Heart Rate Variability and SC increments during intense arousal. | Valence, arousal, dominance<br><br>ECG, GSR, Respiration |
| Shumailov and Gunes [102] | Games | A virtual egg catching game (Egg time game), an exploration of the Earth (Google Earth game), a fight against skeletons by speech and virtual swords (Spell Fighter game), an exploration of artworks (Museum of Fine Arts game), a game to slice fruits by swords (Fruit Ninja game) | F1 score of valence: 0.85, arousal: 0.91 | Valence, arousal<br><br>EMG |
| Ishaque, et al. [113] | Games | Relax- A fish game<br>Stress - Roller coaster experience, Colour Stroop test | Accuracy of 85% for stress classification | ECG, GSR, Respiration<br><br>Stress, relaxation |
| Suhaimi, et al. [77] | 360º videos | Content from YouTube, Within, Discovery VR, Jaunt VR, NYT VR, Veer VR and Google Cardboard based on love, sky drive, animals, and tree climbing, ghost and bloody, dark spirits attacking player | Revealed self-reported discrete emotions | 16 discrete emotions |
| Nazmi Sofian Suhaimi [114] | 360º videos | Arousal and valence space-based content | Accuracy of 85.01% for four-class (calm, bored, angry, happy) classification from EEG signals. | Calm, bored, angry, happy<br><br>EEG |
| Gutiérrez-Maldonado, et al. [21] | Avatars | Avatar faces eliciting happy, sad, fear, anger, disgust, and neutral emotions | Happiness was the top rate of identifying.<br>Virtual faces were misleading in the recognizing of anger with disgust. | Happiness, sadness, fear, disgust, anger, neutral |
| Toet, et al. [103] | 360º videos | Content[43] from previous research by Li, et al. [91] | EmojiGrid tool showed similar results as the previous studies and EmojiGrid related to physiological arousal. | Valence, arousal<br><br>EDA |
| Gupta, et al. [101] | 360º videos | Not given | Accuracy of personalized model: 96.5%, generalized model: 83.7% | Valence<br><br>EEG, GSR |
| Marín-Morales, et al. [92] | 360º panoramas | Architectural scenarios | Accuracy of arousal: 75%, valence: 71.21% | Valence, arousal |
| Suhaimi, et al. [115] | 360º videos | Valence and arousal videos triggering the 4 quadrants of the dimensional model | Accuracy of GBM: 67.04%, NB: 36.24%, RF: 82.49% | Happy, anger, bored, calm<br><br>EEG |
| Teo and Chia [99] | 360º videos | A roller coaster experience with rapid drops, peaks, and speedier turns. | Accuracy of DNN 96.32% | Arousal |

## 12 References

[1] K. R. Scherer, "Measuring the meaning of emotion words: A domain-specific componential approach1," in *Components of Emotional Meaning*Oxford: Oxford University Press, 2013.

[2] C. E. Izard, "Emotion theory and research: highlights, unanswered questions, and emerging issues," (in eng), *Annual review of psychology,* vol. 60, pp. 1-25, 2009.

[3] S. Koelstra *et al.*, "DEAP: A Database for Emotion Analysis ;Using Physiological Signals," *IEEE Transactions on Affective Computing,* vol. 3, no. 1, pp. 18-31, 2012.

[4] L. Shu *et al.*, "A Review of Emotion Recognition Using Physiological Signals," *Sensors,* vol. 18, p. 2074, 06/28 2018.

[5] X. Jia *et al.*, "Multi-Channel EEG Based Emotion Recognition Using Temporal Convolutional Network and Broad Learning System," in *2020 IEEE International Conference on Systems, Man, and Cybernetics (SMC)*, 2020, pp. 2452-2457.

[6] R. W. Picard, "Affective Computing," *Affective Computing,* no. 321.

[7] D. Grandjean and K. Scherer, "Unpacking the Cognitive Architecture of Emotion Processes," *Emotion (Washington, D.C.),* vol. 8, pp. 341-51, 07/01 2008.

[8] J. Leitão, B. Meuleman, D. Van De Ville, and P. Vuilleumier, "Computational imaging during video game playing shows dynamic synchronization of cortical and subcortical networks of emotions," *PLOS Biology,* vol. 18, no. 11, p. e3000900, 2020.

[9] G. Keren, T. Kirschstein, E. Marchi, F. Ringeval, and B. Schuller, "End-to-end learning for dimensional emotion recognition from physiological signals," in *2017 IEEE International Conference on Multimedia and Expo (ICME)*, 2017, pp. 985-990.

[10] S. Alhagry, A. Aly, and R. El-Khoribi, "Emotion Recognition based on EEG using LSTM Recurrent Neural Network," *International Journal of Advanced Computer Science and Applications,* vol. 8, 10/01 2017.

[11] S. Ojha, J. Vitale, and M.-A. Williams, "Computational Emotion Models: A Thematic Review," *International Journal of Social Robotics,* 2020/11/05 2020.

[12] W. Zheng, W. Liu, Y. Lu, B. Lu, and A. Cichocki, "EmotionMeter: A Multimodal Framework for Recognizing Human Emotions," *IEEE Transactions on Cybernetics,* vol. 49, no. 3, pp. 1110-1122, 2019.

[43] https://www.stanfordvr.com/360-video-database/

[13] G. Mohammadi, D. Van De Ville, and P. Vuilleumier, "Brain networks subserving functional core processes of emotions identified with componential modelling," *bioRxiv*, p. 2020.06.10.145201, 2020.
[14] A. Paiva, I. Leite, and T. Ribeiro, "Emotion Modelling for Social Robots," 2014.
[15] E. S. Dan-Glauser and K. R. Scherer, "The Geneva affective picture database (GAPED): a new 730-picture database focusing on valence and normative significance," *Behavior Research Methods*, vol. 43, no. 2, p. 468, 2011/03/24 2011.
[16] J. Kory and S. K. D'Mello, "Affect Elicitation for Affective Computing," 2015.
[17] TED2015, "How virtual reality can create the ultimate empathy machine," ed, 2015, p. https://www.ted.com/talks/chris_milk_how_virtual_reality_can_create_the_ultimate_empathy_machine?language=en.
[18] B. Meuleman and D. Rudrauf, "Induction and profiling of strong multi-componential emotions in virtual reality," *IEEE Transactions on Affective Computing*, vol. PP, pp. 1-1, 08/10 2018.
[19] M. Granato, D. Gadia, D. Maggiorini, and L. A. Ripamonti, "An empirical study of players' emotions in VR racing games based on a dataset of physiological data," *Multimedia Tools and Applications*, 2020/03/02 2020.
[20] X. Peng, J. Huang, A. Denisova, H. Chen, F. Tian, and H. Wang, "A Palette of Deepened Emotions: Exploring Emotional Challenge in Virtual Reality Games," presented at the Proceedings of the 2020 CHI Conference on Human Factors in Computing Systems, Honolulu, HI, USA, 2020. Available: https://doi.org/10.1145/3313831.3376221
[21] J. Gutiérrez-Maldonado, M. Rus-Calafell, and J. González-Conde, "Creation of a new set of dynamic virtual reality faces for the assessment and training of facial emotion recognition ability," *Virtual Reality*, vol. 18, no. 1, pp. 61-71, 2014/03/01 2014.
[22] S. Estupiñán, F. Rebelo, P. Noriega, C. Ferreira, and E. Duarte, "Can Virtual Reality Increase Emotional Responses (Arousal and Valence)? A Pilot Study," Cham, 2014, pp. 541-549: Springer International Publishing.
[23] P. J. Bota, C. Wang, A. L. N. Fred, and H. P. D. Silva, "A Review, Current Challenges, and Future Possibilities on Emotion Recognition Using Machine Learning and Physiological Signals," *IEEE Access*, vol. 7, pp. 140990-141020, 2019.
[24] E. Siedlecka and T. F. Denson, "Experimental Methods for Inducing Basic Emotions: A Qualitative Review," vol. 11, no. 1, pp. 87-97, 2019.
[25] G. Riva *et al.*, "Affective Interactions Using Virtual Reality: The Link between Presence and Emotions," *Cyberpsychology & behavior : the impact of the Internet, multimedia and virtual reality on behavior and society*, vol. 10, pp. 45-56, 03/01 2007.
[26] A. Kim, M. Chang, Y. Choi, S. Jeon, and K. Lee, *The Effect of Immersion on Emotional Responses to Film Viewing in a Virtual Environment*. 2018, pp. 601-602.
[27] H. Lench, S. Flores, and S. Bench, "Discrete Emotions Predict Changes in Cognition, Judgment, Experience, Behavior, and Physiology: A Meta-Analysis of Experimental Emotion Elicitations," *Psychological bulletin*, vol. 137, pp. 834-55, 09/01 2011.
[28] E. Harmon-Jones, C. Harmon-Jones, and E. Summerell, "On the Importance of Both Dimensional and Discrete Models of Emotion," (in eng), *Behav Sci (Basel)*, vol. 7, no. 4, Sep 29 2017.
[29] J. J. R. Fontaine, "Dimensional, basic emotion, and componential approaches to meaning in psychological emotion research1," in *Components of Emotional Meaning*Oxford: Oxford University Press, 2013.
[30] D. Grandjean, D. Sander, and K. Scherer, "Conscious emotional experience emerges as a function of multilevel, appraisal-driven response synchronization," *Consciousness and cognition*, vol. 17, pp. 484-95, 07/01 2008.
[31] P. Ekman, "Are there basic emotions?," *Psychological Review*, vol. 99, no. 3, pp. 550-553, 1992.
[32] S. Gu, F. Wang, N. P. Patel, J. A. Bourgeois, and J. H. Huang, "A Model for Basic Emotions Using Observations of Behavior in Drosophila," (in English), Review vol. 10, no. 781, 2019-April-24 2019.
[33] P. Ekman, "Basic Emotions," in *Handbook of Cognition and Emotion*, 2005, pp. 45-60.

[34] C. E. Izard, "Basic Emotions, Natural Kinds, Emotion Schemas, and a New Paradigm," vol. 2, no. 3, pp. 260-280, 2007.

[35] M. Egger, M. Ley, and S. Hanke, "Emotion Recognition from Physiological Signal Analysis: A Review," *Electronic Notes in Theoretical Computer Science,* vol. 343, pp. 35-55, 2019/05/04/ 2019.

[36] J. A. Domínguez-Jiménez, K. C. Campo-Landines, J. C. Martínez-Santos, E. J. Delahoz, and S. H. Contreras-Ortiz, "A machine learning model for emotion recognition from physiological signals," *Biomedical Signal Processing and Control,* vol. 55, p. 101646, 2020/01/01/ 2020.

[37] P. Ekman, W. J. J. o. p. Friesen, and s. psychology, "Constants across cultures in the face and emotion," vol. 17 2, pp. 124-9, 1971.

[38] W. G. Parrott, *Emotions in Social Psychology: Essential Readings*. Psychology Press, 2001.

[39] N. H. Frijda, N. H. A. FRIDJA, A. Manstead, and K. Oatley, *The Emotions*. Cambridge University Press, 1986.

[40] R. Jack, O. Garrod, and P. Schyns, "Dynamic Facial Expressions of Emotion Transmit an Evolving Hierarchy of Signals over Time," *Current biology : CB,* vol. 24, 12/31 2013.

[41] S. Gu, F. Wang, T. Yuan, B. Guo, and J. H. J. I. j. o. n. Huang, "Differentiation of Primary Emotions through Neuromodulators: Review of Literature," vol. 1, pp. 43-50, 2015.

[42] A. Pereira Junior and F. Wang, "Neuromodulation, Emotional Feelings and Affective Disorders," *Mens Sana Monographs,* vol. 0, 10/26 2016.

[43] Z. Zheng *et al.*, "Safety Needs Mediate Stressful Events Induced Mental Disorders," *Neural Plasticity,* vol. 2016, pp. 1-6, 01/01 2016.

[44] M. T. Cicero and M. Graver, *Cicero on the Emotions: Tusculan Disputations 3 and 4*. University of Chicago Press, 2002.

[45] J. Russell, "A Circumplex Model of Affect," *Journal of Personality and Social Psychology,* vol. 39, pp. 1161-1178, 12/01 1980.

[46] J. Posner, J. A. Russell, and B. S. Peterson, "The circumplex model of affect: an integrative approach to affective neuroscience, cognitive development, and psychopathology," (in eng), *Development and psychopathology,* vol. 17, no. 3, pp. 715-734, Summer 2005.

[47] H. Schlosberg, "Three dimensions of emotion," vol. 61, ed. US: American Psychological Association, 1954, pp. 81-88.

[48] P. Ekman, "A methodological discussion of nonverbal behavior," *The Journal of Psychology: Interdisciplinary and Applied,* vol. 43, pp. 141-149, 1957.

[49] W. M. Wundt, *Principles of Physiological Psychology*. Kraus Reprint, 1969.

[50] C. E. Osgood, W. H. May, M. S. Miron, and M. S. Miron, *Cross-cultural Universals of Affective Meaning* (no. v. 1). University of Illinois Press, 1975.

[51] G. Mohammadi and P. Vuilleumier, "A Multi-Componential Approach to Emotion Recognition and the Effect of Personality," *IEEE Transactions on Affective Computing,* pp. 1-1, 2020.

[52] A. Moors, P. Ellsworth, K. Scherer, and N. Frijda, "Appraisal Theories of Emotion: State of the Art and Future Development," *Emotion Review,* vol. 5, pp. 119-124, 03/20 2013.

[53] K. R. Scherer, "The dynamic architecture of emotion: Evidence for the component process model," *Cognition and Emotion,* vol. 23, no. 7, pp. 1307-1351, 2009/11/01 2009.

[54] A. Marchewka, Ł. Żurawski, K. Jednoróg, and A. Grabowska, "The Nencki Affective Picture System (NAPS): Introduction to a novel, standardized, wide-range, high-quality, realistic picture database," *Behavior Research Methods,* vol. 46, no. 2, pp. 596-610, 2014/06/01 2014.

[55] S. D. Kreibig, "Autonomic nervous system activity in emotion: A review," *Biological Psychology,* vol. 84, no. 3, pp. 394-421, 2010/07/01/ 2010.

[56] L. Hinkle, K. Khoshhal, and V. Metsis, "Physiological Measurement for Emotion Recognition in Virtual Reality," in *2019 2nd International Conference on Data Intelligence and Security (ICDIS)*, 2019, pp. 136-143.

[57] E. Bekele, D. Bian, J. Peterman, S. Park, and N. Sarkar, "Design of a Virtual Reality System for Affect Analysis in Facial Expressions (VR-SAAFE); Application to Schizophrenia," *IEEE Transactions on Neural Systems and Rehabilitation Engineering,* vol. 25, no. 6, pp. 739-749, 2017.

[58] O. Bălan *et al.*, "Sensors system methodology for artefacts identification in Virtual Reality games," in *2019 International Symposium on Advanced Electrical and Communication Technologies (ISAECT)*, 2019, pp. 1-6.
[59] P. J. Lang, "International Affective Picture System (IAPS) : Technical Manual and Affective Ratings," 1995.
[60] W. Yang *et al.*, "Affective auditory stimulus database: An expanded version of the International Affective Digitized Sounds (IADS-E)," *Behavior Research Methods*, vol. 50, no. 4, pp. 1415-1429, 2018/08/01 2018.
[61] A. AlzeerAlhouseini, I. Al-Shaikhli, A. AbdulRahman, and M. J. I. J. o. A. i. C. T. Dzulkifli, "Emotion detection using physiological signals EEG & ECG," vol. 8, no. 3, pp. 103-112, 2016.
[62] S. Katsigiannis and N. Ramzan, "DREAMER: A Database for Emotion Recognition Through EEG and ECG Signals From Wireless Low-cost Off-the-Shelf Devices," *IEEE Journal of Biomedical and Health Informatics*, vol. 22, no. 1, pp. 98-107, 2018.
[63] G. Mohammadi, K. Lin, and P. Vuilleumier, "Towards Understanding Emotional Experience in a Componential Framework," in *2019 8th International Conference on Affective Computing and Intelligent Interaction (ACII)*, 2019, pp. 123-129.
[64] R. Subramanian, J. Wache, M. K. Abadi, R. L. Vieriu, S. Winkler, and N. Sebe, "ASCERTAIN: Emotion and Personality Recognition Using Commercial Sensors," *IEEE Transactions on Affective Computing*, vol. 9, no. 2, pp. 147-160, 2018.
[65] J. A. M. Correa, M. K. Abadi, N. Sebe, and I. Patras, "AMIGOS: A Dataset for Affect, Personality and Mood Research on Individuals and Groups," *IEEE Transactions on Affective Computing*, pp. 1-1, 2018.
[66] K. Hidaka, H. Qin, and J. Kobayashi, "Preliminary test of affective virtual reality scenes with head mount display for emotion elicitation experiment," in *2017 17th International Conference on Control, Automation and Systems (ICCAS)*, 2017, pp. 325-329.
[67] W. Zhang, L. Shu, X. Xu, and D. Liao, "Affective Virtual Reality System (AVRS): Design and Ratings of Affective VR Scenes," in *2017 International Conference on Virtual Reality and Visualization (ICVRV)*, 2017, pp. 311-314.
[68] T. Xu, R. Yin, L. Shu, and X. Xu, "Emotion Recognition Using Frontal EEG in VR Affective Scenes," in *2019 IEEE MTT-S International Microwave Biomedical Conference (IMBioC)*, 2019, vol. 1, pp. 1-4.
[69] S. Susindar, M. Sadeghi, L. Huntington, A. Singer, and T. K. Ferris, "The Feeling is Real: Emotion Elicitation in Virtual Reality," vol. 63, no. 1, pp. 252-256, 2019.
[70] B. U. C. Simo Järvelä, Mikko Salminen, Giulio Jacucci, Juho Hamari, Niklas Ravaja, "Augmented Virtual Reality Meditation: Shared Dyadic Biofeedback Increases Social Presence Via Respiratory Synchrony," *ACM Transactions on Social Computing*, 2021.
[71] M. El-Yamri, A. Romero-Hernandez, M. Gonzalez-Riojo, and B. Manero, "Emotions-Responsive Audiences for VR Public Speaking Simulators Based on the Speakers' Voice," in *2019 IEEE 19th International Conference on Advanced Learning Technologies (ICALT)*, 2019, vol. 2161-377X, pp. 349-353.
[72] I.-C. Stanica, M. Dascalu, C. Bodea, and A. Moldoveanu, *VR Job Interview Simulator: Where Virtual Reality Meets Artificial Intelligence for Education*. 2018, pp. 9-12.
[73] C. Rojas, M. Corral, N. Poulsen, and P. Maes, "Project Us: A Wearable for Enhancing Empathy," presented at the Companion Publication of the 2020 ACM Designing Interactive Systems Conference, Eindhoven, Netherlands, 2020. Available: https://doi.org/10.1145/3393914.3395882
[74] A. Chirico, D. B. Yaden, G. Riva, and A. Gaggioli, "The Potential of Virtual Reality for the Investigation of Awe," (in eng), *Front Psychol*, vol. 7, p. 1766, 2016.
[75] F. Pallavicini and A. Pepe, "Virtual Reality Games and the Role of Body Involvement in Enhancing Positive Emotions and Decreasing Anxiety: Within-Subjects Pilot Study," (in eng), *JMIR Serious Games*, vol. 8, no. 2, p. e15635, Jun 17 2020.
[76] T. B. Alakus, M. Gonen, and I. Turkoglu, "Database for an emotion recognition system based on EEG signals and various computer games – GAMEEMO," *Biomedical Signal Processing and Control*, vol. 60, p. 101951, 2020/07/01/ 2020.

[77] N. S. Suhaimi, C. T. B. Yuan, J. Teo, and J. Mountstephens, "Modeling the affective space of 360 virtual reality videos based on arousal and valence for wearable EEG-based VR emotion classification," in *2018 IEEE 14th International Colloquium on Signal Processing & Its Applications (CSPA)*, 2018, pp. 167-172.

[78] Y. Li, A. S. Elmaghraby, and E. M. Sokhadze, "Designing immersive affective environments with biofeedback," in *2015 Computer Games: AI, Animation, Mobile, Multimedia, Educational and Serious Games (CGAMES)*, 2015, pp. 73-77.

[79] I. G. Amit Barde, Ashkan F. Hayati, Arindam Dey, Gun Lee, Mark Billinghurst, "A Review of Hyperscanning and Its Use in Virtual Environments," *Informatics*, vol. 7, no. 4, 2020.

[80] J.-H. T. Lin, "Fear in virtual reality (VR): Fear elements, coping reactions, immediate and next-day fright responses toward a survival horror zombie virtual reality game," *Computers in Human Behavior*, vol. 72, pp. 350-361, 2017/07/01/ 2017.

[81] K. Gupta, R. Hajika, Y. S. Pai, A. Duenser, M. Lochner, and M. Billinghurst, "Measuring Human Trust in a Virtual Assistant using Physiological Sensing in Virtual Reality," in *2020 IEEE Conference on Virtual Reality and 3D User Interfaces (VR)*, 2020, pp. 756-765.

[82] B. Cebeci, U. Celikcan, and T. K. Capin, "A comprehensive study of the affective and physiological responses induced by dynamic virtual reality environments," vol. 30, no. 3-4, p. e1893, 2019.

[83] O. I. Caldas, O. F. Aviles, and C. Rodriguez-Guerrero, "Effects of Presence and Challenge Variations on Emotional Engagement in Immersive Virtual Environments," *IEEE Transactions on Neural Systems and Rehabilitation Engineering*, vol. 28, no. 5, pp. 1109-1116, 2020.

[84] S. Michalski, A. Szpak, D. Saredakis, T. Ross, M. Billinghurst, and T. Loetscher, "Getting your game on: Using virtual reality to improve real table tennis skills," *PLOS ONE*, vol. 14, p. e0222351, 09/10 2019.

[85] L. Reidy, D. Chan, C. Nduka, and H. J. a. p. a. Gunes, "Facial Electromyography-based Adaptive Virtual Reality Gaming for Cognitive Training," 2020.

[86] L. McMichael *et al.*, "Parents of Adolescents Perspectives of Physical Activity, Gaming and Virtual Reality: Qualitative Study," *JMIR Serious Games*, 08/25 2020.

[87] S. Lv, Q. Zhang, and L. Wang, "VR virtual reality technology and treatment progress," *SID Symposium Digest of Technical Papers*, vol. 51, pp. 35-38, 07/01 2020.

[88] E. L. M. Naves, T. F. Bastos, G. Bourhis, Y. M. L. R. Silva, V. J. Silva, and V. F. Lucena, "Virtual and augmented reality environment for remote training of wheelchairs users: Social, mobile, and wearable technologies applied to rehabilitation," in *2016 IEEE 18th International Conference on e-Health Networking, Applications and Services (Healthcom)*, 2016, pp. 1-4.

[89] K. Rahul, V. P. D. Raj, K. Srinivasan, N. Deepa, and N. S. Kumar, "A Study on Virtual and Augmented Reality in Real-Time Surgery," in *2019 IEEE International Conference on Consumer Electronics - Taiwan (ICCE-TW)*, 2019, pp. 1-2.

[90] J. Nam, H. Chung, Y. a. Seong, and H. J. A. Lee, "A New Terrain in HCI: Emotion Recognition Interface using Biometric Data for an Immersive VR Experience," vol. abs/1912.01177, 2019.

[91] B. J. Li, J. N. Bailenson, A. Pines, W. J. Greenleaf, and L. M. Williams, "A Public Database of Immersive VR Videos with Corresponding Ratings of Arousal, Valence, and Correlations between Head Movements and Self Report Measures," (in English), Original Research vol. 8, no. 2116, 2017-December-05 2017.

[92] J. Marín-Morales *et al.*, "Affective computing in virtual reality: emotion recognition from brain and heartbeat dynamics using wearable sensors," *Scientific Reports*, vol. 8, no. 1, p. 13657, 2018/09/12 2018.

[93] T. Teo, M. Norman, G. Lee, M. Billinghurst, and M. Adcock, "Exploring interaction techniques for 360 panoramas inside a 3D reconstructed scene for mixed reality remote collaboration," *Journal on Multimodal User Interfaces*, 07/25 2020.

[94] T. Teo, L. Lawrence, G. A. Lee, M. Billinghurst, and M. Adcock, "Mixed Reality Remote Collaboration Combining 360 Video and 3D Reconstruction," presented at the Proceedings of the 2019 CHI Conference on Human Factors in Computing Systems, Glasgow, Scotland Uk, 2019. Available: https://doi.org/10.1145/3290605.3300431

[95] D. Liao *et al.*, "Arousal Evaluation of VR Affective Scenes Based on HR and SAM," in *2019 IEEE MTT-S International Microwave Biomedical Conference (IMBioC)*, 2019, vol. 1, pp. 1-4.
[96] S. M. Hofmann, F. Klotzsche, A. Mariola, V. V. Nikulin, A. Villringer, and M. Gaebler, "Decoding Subjective Emotional Arousal during a Naturalistic VR Experience from EEG Using LSTMs," in *2018 IEEE International Conference on Artificial Intelligence and Virtual Reality (AIVR)*, 2018, pp. 128-131.
[97] I. Mavridou, E. Seiss, T. Kostoulas, C. Nduka, and E. Balaguer-Ballester, "Towards an effective arousal detection system for virtual reality," presented at the Proceedings of the Workshop on Human-Habitat for Health (H3): Human-Habitat Multimodal Interaction for Promoting Health and Well-Being in the Internet of Things Era, Boulder, Colorado, 2018. Available: https://doi.org/10.1145/3279963.3279969
[98] Y. Ding *et al.*, "TSception:A Deep Learning Framework for Emotion Detection Using EEG," pp. 1-7, 2020.
[99] J. Teo and J. T. Chia, "Deep Neural Classifiers For Eeg-Based Emotion Recognition In Immersive Environments," in *2018 International Conference on Smart Computing and Electronic Enterprise (ICSCEE)*, 2018, pp. 1-6.
[100] I. Mavridou, E. Seiss, M. Hamedi, E. Balaguer-Ballester, and C. Nduka, "Towards valence detection from EMG for Virtual Reality applications," presented at the 12th International Conference on Disability, Virtual Reality and Associated Technologies (ICDVRAT 2018), Nottingham, UK, 2018. Available: http://eprints.bournemouth.ac.uk/31022/
[101] K. Gupta, J. Lazarevic, Y. S. Pai, and M. Billinghurst, "AffectivelyVR: Towards VR Personalized Emotion Recognition," presented at the 26th ACM Symposium on Virtual Reality Software and Technology, Virtual Event, Canada, 2020. Available: https://doi.org/10.1145/3385956.3422122
[102] I. Shumailov and H. Gunes, "Computational analysis of valence and arousal in virtual reality gaming using lower arm electromyograms," in *2017 Seventh International Conference on Affective Computing and Intelligent Interaction (ACII)*, 2017, pp. 164-169.
[103] A. Toet, F. Heijn, A.-M. Brouwer, T. Mioch, and J. B. F. van Erp, "An Immersive Self-Report Tool for the Affective Appraisal of 360° VR Videos," (in English), Original Research vol. 1, no. 14, 2020-September-25 2020.
[104] A. Dey, H. Chen, M. Billinghurst, and R. W. Lindeman, "Effects of Manipulating Physiological Feedback in Immersive Virtual Environments," presented at the Proceedings of the 2018 Annual Symposium on Computer-Human Interaction in Play, Melbourne, VIC, Australia, 2018. Available: https://doi.org/10.1145/3242671.3242676
[105] M. Moghimi, R. Stone, and P. Rotshtein, "Affective Recognition in Dynamic and Interactive Virtual Environments," *IEEE Transactions on Affective Computing*, vol. 11, no. 1, pp. 45-62, 2020.
[106] A. Felnhofer *et al.*, "Is virtual reality emotionally arousing? Investigating five emotion inducing virtual park scenarios," *International Journal of Human-Computer Studies*, vol. 82, pp. 48-56, 2015/10/01/ 2015.
[107] C. Bassano *et al.*, "A VR Game-based System for Multimodal Emotion Data Collection," presented at the Motion, Interaction and Games, Newcastle upon Tyne, United Kingdom, 2019. Available: https://doi.org/10.1145/3359566.3364695
[108] G. Wilson and M. McGill, "Violent Video Games in Virtual Reality: Re-Evaluating the Impact and Rating of Interactive Experiences," in *Proceedings of the 2018 Annual Symposium on Computer-Human Interaction in Play*: Association for Computing Machinery, 2018, pp. 535–548.
[109] F. Pallavicini, A. Pepe, and M. E. Minissi, "Gaming in Virtual Reality: What Changes in Terms of Usability, Emotional Response and Sense of Presence Compared to Non-Immersive Video Games?," *Simulation & Gaming*, vol. 50, p. 104687811983142, 03/05 2019.
[110] A. Chirico, F. Ferrise, L. Cordella, and A. Gaggioli, "Designing Awe in Virtual Reality: An Experimental Study," (in English), Original Research vol. 8, no. 2351, 2018-January-22 2018.
[111] N. Lipp, N. Dużmańska-Misiarczyk, A. Strojny, and P. Strojny, "Evoking emotions in virtual reality: schema activation via a freeze-frame stimulus," *Virtual Reality*, 2020/07/03 2020.

[112] S. C. Steinhaeusser and B. Lugrin, "Horror Laboratory and Forest Cabin - A Horror Game Series for Desktop Computer, Virtual Reality, and Smart Substitutional Reality," presented at the Extended Abstracts of the 2020 Annual Symposium on Computer-Human Interaction in Play, Virtual Event, Canada, 2020. Available: https://doi.org/10.1145/3383668.3419924
[113] S. Ishaque, A. Rueda, B. Nguyen, N. Khan, and S. Krishnan, "Physiological Signal Analysis and Classification of Stress from Virtual Reality Video Game," in *2020 42nd Annual International Conference of the IEEE Engineering in Medicine & Biology Society (EMBC)*, 2020, pp. 867-870.
[114] J. M. J. T. Nazmi Sofian Suhaimi, "Parameter Tuning for Enhancing Inter-Subject Emotion Classification in Four Classes for VR-EEG Predictive Analytics," *International Journal of Advanced Science and Technology*, vol. 29, no. 6s, pp. 1483 - 1491, 04/14 2020.
[115] N. S. B. Suhaimi, J. Mountstephens, and J. Teo, "Emotional State Classification with Distributed Random Forest, Gradient Boosting Machine and Naïve Bayes in Virtual Reality Using Wearable Electroencephalography and Inertial Sensing," in *2020 IEEE 10th Symposium on Computer Applications & Industrial Electronics (ISCAIE)*, 2020, pp. 12-17.
[116] I. Mavridou, E. Seiss, T. Kostoulas, M. Hamedi, E. Balaguer-Ballester, and C. Nduka, "Introducing the EmteqVR Interface for Affect Detection in Virtual Reality," in *2019 8th International Conference on Affective Computing and Intelligent Interaction Workshops and Demos (ACIIW)*, 2019, pp. 83-84.
[117] D. Smeijers and S. L. Koole, "Testing the Effects of a Virtual Reality Game for Aggressive Impulse Management (VR-GAIME): Study Protocol," (in English), Clinical Study Protocol vol. 10, no. 83, 2019-February-26 2019.
[118] N. Reski and A. Alissandrakis, "Open data exploration in virtual reality: a comparative study of input technology," *Virtual Reality*, vol. 24, no. 1, pp. 1-22, 2020/03/01 2020.
[119] C. de-Juan-Ripoll, J. Llanes-Jurado, I. A. C. Giglioli, J. Marín-Morales, and M. Alcañiz, "An Immersive Virtual Reality Game for Predicting Risk Taking through the Use of Implicit Measures," vol. 11, no. 2, p. 825, 2021.
[120] A. Dey, J. Phoon, S. Saha, C. Dobbins, and M. Billinghurst, "A Neurophysiological Approach for Measuring Presence in Immersive Virtual Environments," in *2020 IEEE International Symposium on Mixed and Augmented Reality (ISMAR)*, 2020, pp. 474-485.
[121] Y. Kano and J. Morita, "The Effect of Experience and Embodiment on Empathetic Behavior toward Virtual Agents," presented at the Proceedings of the 8th International Conference on Human-Agent Interaction, Virtual Event, USA, 2020. Available: https://doi.org/10.1145/3406499.3415074
[122] Y. Li, A. S. Elmaghraby, A. El-Baz, and E. M. Sokhadze, "Using physiological signal analysis to design affective VR games," in *2015 IEEE International Symposium on Signal Processing and Information Technology (ISSPIT)*, 2015, pp. 57-62.
[123] Z. Zhao, F. Han, and X. Ma, "A Live Storytelling Virtual Reality System with Programmable Cartoon-Style Emotion Embodiment," in *2019 IEEE International Conference on Artificial Intelligence and Virtual Reality (AIVR)*, 2019, pp. 102-1027.
[124] S. Oberdörfer, D. Heidrich, and M. E. Latoschik, "Think Twice: The Influence of Immersion on Decision Making during Gambling in Virtual Reality," in *2020 IEEE Conference on Virtual Reality and 3D User Interfaces (VR)*, 2020, pp. 483-492.
[125] N. G. Arvind Kumar, Gurpreet Kaur, "An Emotion Recognition Based on Physiological Signals," *International Journal of Innovative Technology and Exploring Engineering (IJITEE)*, vol. 8, no. 9S, July 2019 2019.
[126] K. Gouizi, F. Bereksi Reguig, and C. Maaoui, "Emotion recognition from physiological signals," *Journal of Medical Engineering & Technology*, vol. 35, no. 6-7, pp. 300-307, 2011/10/01 2011.
[127] M. Horvat, M. Dobrinić, M. Novosel, and P. Jerčić, *Assessing emotional responses induced in virtual reality using a consumer EEG headset: A preliminary report*. 2018.
[128] P. Bilgin, K. Agres, N. Robinson, A. A. P. Wai, and C. Guan, "A Comparative Study of Mental States in 2D and 3D Virtual Environments Using EEG," in *2019 IEEE International Conference on Systems, Man and Cybernetics (SMC)*, 2019, pp. 2833-2838.

[129] Y. Sekhavat, S. Roohi, H. Sakian, and G. Yannakakis, *Play with One's Feelings: A Study on Emotion Awareness for Player Experience*. 2020.

[130] D. Weibel and B. Wissmath, "Immersion in Computer Games: The Role of Spatial Presence and Flow," *International Journal of Computer Games Technology,* vol. 2011, p. 282345, 2012/01/05 2011.

[131] H. P. Martinez, Y. Bengio, and G. N. Yannakakis, "Learning deep physiological models of affect," *IEEE Computational Intelligence Magazine,* vol. 8, no. 2, pp. 20-33, 2013.

[132] C. van Reekum, T. Johnstone, R. Banse, A. Etter, T. Wehrle, and K. Scherer, "Psychophysiological responses to appraisal dimensions in a computer game," *Cognition and Emotion,* vol. 18, no. 5, pp. 663-688, 2004/08/01 2004.

[133] J. M. J. T. Aaron Frederick Bulagang, "Four-Class Emotion Classification using Electrocardiography (ECG) in Virtual Reality (VR)," *International Journal of Advanced Science and Technology,* vol. 29, no. 6s, pp. 1523 - 1529, 04/14 2020.

[134] L. J. Zheng, J. Mountstephens, and J. Teo, "Four-class emotion classification in virtual reality using pupillometry," *Journal of Big Data,* vol. 7, no. 1, p. 43, 2020/07/06 2020.

[135] M. Soleymani, J. Lichtenauer, T. Pun, and M. Pantic, "A Multimodal Database for Affect Recognition and Implicit Tagging," *IEEE Transactions on Affective Computing,* vol. 3, no. 1, pp. 42-55, 2012.

[136] J. I. Vallade, R. Kaufmann, B. N. Frisby, and J. C. Martin, "Technology acceptance model: investigating students' intentions toward adoption of immersive 360° videos for public speaking rehearsals," *Communication Education,* pp. 1-19, 2020.

[137] M. Feidakis, "Chapter 11 - A Review of Emotion-Aware Systems for e-Learning in Virtual Environments," in *Formative Assessment, Learning Data Analytics and Gamification*, S. Caballé and R. Clarisó, Eds. Boston: Academic Press, 2016, pp. 217-242.

[138] L. Santamaria-Granados, M. Munoz-Organero, G. Ramirez-González, E. Abdulhay, and N. Arunkumar, "Using Deep Convolutional Neural Network for Emotion Detection on a Physiological Signals Dataset (AMIGOS)," *IEEE Access,* vol. 7, pp. 57-67, 2019.

[139] V. Delvigne, H. Wannous, T. Dutoit, L. Ris, and J. P. Vandeborre, "PhyDAA: Physiological Dataset Assessing Attention," *IEEE Transactions on Circuits and Systems for Video Technology,* pp. 1-1, 2021.

[140] M. M. Bradley and P. J. Lang, "Measuring emotion: The self-assessment manikin and the semantic differential," *Journal of Behavior Therapy and Experimental Psychiatry,* vol. 25, no. 1, pp. 49-59, 1994/03/01/ 1994.

[141] J. Difede and H. G. Hoffman, "Virtual reality exposure therapy for World Trade Center Post-traumatic Stress Disorder: a case report," (in eng), *Cyberpsychol Behav,* vol. 5, no. 6, pp. 529-35, Dec 2002.

[142] V. Shuman, K. Schlegel, and K. Scherer, *Geneva Emotion Wheel Rating Study*. 2015.

[143] V. Tran, "Positive Affect Negative Affect Scale (PANAS)," in *Encyclopedia of Behavioral Medicine*, M. D. Gellman and J. R. Turner, Eds. New York, NY: Springer New York, 2013, pp. 1508-1509.

[144] G. J. J. P. Boyle and I. Differences, "Reliability and validity of Izard's differential emotions scale," vol. 5, no. 6, pp. 747-750, 1984.

[145] M. Schröder, H. Pirker, and M. Lamolle, "First suggestions for an emotion annotation and representation language," in *Proceedings of LREC,* 2006, vol. 6, pp. 88-92.

[146] J. J. Gross and R. W. Levenson, "Emotion elicitation using films," *Cognition and Emotion,* vol. 9, no. 1, pp. 87-108, 1995/01/01 1995.

[147] A. Betella and P. F. M. J. Verschure, "The Affective Slider: A Digital Self-Assessment Scale for the Measurement of Human Emotions," *PLOS ONE,* vol. 11, no. 2, p. e0148037, 2016.

[148] A. F. Bulagang, J. Mountstephens, and J. Teo, "Multiclass emotion prediction using heart rate and virtual reality stimuli," *Journal of Big Data,* vol. 8, no. 1, p. 12, 2021/01/07 2021.

[149] J. J. R. Fontaine, K. R. Scherer, and C. Soriano, "The why, the what, and the how of the GRID instrument1," in *Components of Emotional Meaning*Oxford: Oxford University Press, 2013.

[150] K. R. Scherer, J. R. F. Fontaine, and C. Soriano, "CoreGRID and MiniGRID: Development and validation of two short versions of the GRID instrument1," in *Components of Emotional Meaning*Oxford: Oxford University Press, 2013.

[151] N. Jadhav and R. Sugandhi, "Survey on Human Behavior Recognition Using Affective Computing," in *2018 IEEE Global Conference on Wireless Computing and Networking (GCWCN)*, 2018, pp. 98-103.
[152] R. W. Picard and J. Healey, "Affective wearables," *Personal Technologies*, vol. 1, no. 4, pp. 231-240, 1997/12/01 1997.
[153] M. M. Hassan, M. G. R. Alam, M. Z. Uddin, S. Huda, A. Almogren, and G. Fortino, "Human emotion recognition using deep belief network architecture," *Information Fusion*, vol. 51, pp. 10-18, 2019/11/01/ 2019.
[154] G. Bernal, T. Yang, A. Jain, and P. Maes, "PhysioHMD: a conformable, modular toolkit for collecting physiological data from head-mounted displays," presented at the Proceedings of the 2018 ACM International Symposium on Wearable Computers, Singapore, Singapore, 2018. Available: https://doi.org/10.1145/3267242.3267268
[155] W. Sato, T. Kochiyama, and S. Yoshikawa, "Physiological correlates of subjective emotional valence and arousal dynamics while viewing films," *Biological Psychology*, vol. 157, p. 107974, 2020/11/01/ 2020.
[156] C. Baker, R. Pawling, and S. Fairclough, "Assessment of threat and negativity bias in virtual reality," *Scientific Reports*, vol. 10, no. 1, p. 17338, 2020/10/15 2020.
[157] I. Mavridou *et al.*, "FACETEQ: A novel platform for measuring emotion in VR," in *VRIC '17*, 2017.
[158] I. Mavridou *et al.*, "FACETEQ interface demo for emotion expression in VR," in *2017 IEEE Virtual Reality (VR)*, 2017, pp. 441-442.
[159] e. labs, "Measuring what Matters in Immersive Environments," https://www.emteqlabs.com/2020.
[160] A. Jo and B. Y. Chae, "Introduction to real time user interaction in virtual reality powered by brain computer interface technology," presented at the ACM SIGGRAPH 2020 Real-Time Live!, Virtual Event, USA, 2020. Available: https://doi.org/10.1145/3407662.3407754
[161] C. Sá, P. V. Gomes, A. Marques, and A. Correia, "The Use of Portable EEG Devices in Development of Immersive Virtual Reality Environments for Converting Emotional States into Specific Commands," vol. 54, no. 1, p. 43, 2020.
[162] J. Jantz, A. Molnar, and R. Alcaide, "A brain-computer interface for extended reality interfaces," presented at the ACM SIGGRAPH 2017 VR Village, Los Angeles, California, 2017. Available: https://doi.org/10.1145/3089269.3089290
[163] P. Lakhan *et al.*, "Consumer Grade Brain Sensing for Emotion Recognition," *IEEE Sensors Journal*, vol. 19, no. 21, pp. 9896-9907, 2019.
[164] J. Amores, R. Richer, N. Zhao, P. Maes, and B. M. Eskofier, "Promoting relaxation using virtual reality, olfactory interfaces and wearable EEG," in *2018 IEEE 15th International Conference on Wearable and Implantable Body Sensor Networks (BSN)*, 2018, pp. 98-101.
[165] J. F. Golding, "Predicting individual differences in motion sickness susceptibility by questionnaire," *Personality and Individual Differences*, vol. 41, no. 2, pp. 237-248, 2006/07/01/ 2006.
[166] S. Lamb and K. Kwok, "MSSQ-Short Norms May Underestimate Highly Susceptible Individuals," *Human Factors: The Journal of the Human Factors and Ergonomics Society*, vol. 57, 10/01 2014.
[167] H. K. Kim, J. Park, Y. Choi, and M. Choe, "Virtual reality sickness questionnaire (VRSQ): Motion sickness measurement index in a virtual reality environment," *Applied Ergonomics*, vol. 69, pp. 66-73, 2018/05/01/ 2018.
[168] Y. Konishi, N. Hanamitsu, K. Minamizawa, A. Sato, and T. Mizuguchi, *Synesthesia suit: the full body immersive experience*. 2016, pp. 1-1.
[169] L. F. Barrett, "Solving the emotion paradox: categorization and the experience of emotion," (in eng), *Pers Soc Psychol Rev*, vol. 10, no. 1, pp. 20-46, 2006.
[170] E. Malbos, G. H. Burgess, and C. Lançon, "Virtual Reality and Fear of Shark Attack: A Case Study for the Treatment of Squalophobia," vol. 0, no. 0, p. 1534650120940014.
[171] J. Marín-Morales, J. L. Higuera-Trujillo, C. Llinares, J. Guixeres, M. Alcañiz, and G. Valenza, "Real vs. Immersive Virtual Emotional Museum Experience: a Heart Rate Variability Analysis

during a Free Exploration Task," in *2020 11th Conference of the European Study Group on Cardiovascular Oscillations (ESGCO)*, 2020, pp. 1-2.

[172] J. Myung and H. Jun, "Emotional responses to virtual reality-based 3D spaces: focusing on ECG Response to single-person housing according to different plan configurations," *Journal of Asian Architecture and Building Engineering*, pp. 1-13, 2020.

[173] M. Gnacek *et al.*, "Heart rate detection from the supratrochlear vessels using a virtual reality headset integrated PPG sensor," presented at the 1st International Workshop on Multimodal Affect and Aesthetic Experience on ICMI 2020: 22nd ACM International Conference on Multimodal Interaction, Utrecht, the Netherlands, 2020. Available: http://eprints.bournemouth.ac.uk/34761/

[174] S. Schmidt, O. Ariza, and F. Steinicke, "Intelligent Blended Agents: Reality–Virtuality Interaction with Artificially Intelligent Embodied Virtual Humans," *Multimodal Technologies and Interaction*, vol. 4, p. 85, 11/27 2020.

[175] K. M. R. Thejaswini S, "Electroencephalogram Based Emotion Detection Using Hybrid Long Short Term Memory," *European Journal of Molecular & Clinical Medicine*, vol. 07, no. 08, 2020.

[176] J. D. Hart, T. Piumsomboon, L. Lawrence, G. A. Lee, R. T. Smith, and M. Billinghurst, "Demonstrating Emotion Sharing and Augmentation in Cooperative Virtual Reality Games," in *2018 IEEE International Symposium on Mixed and Augmented Reality Adjunct (ISMAR-Adjunct)*, 2018, pp. 405-406.

[177] A. Kim, M. W. Kim, H. Bae, and K.-M. Lee, "Exploring the Relative Effects of Body Position and Spatial Cognition on Presence When Playing Virtual Reality Games," *International Journal of Human–Computer Interaction*, vol. 36, no. 18, pp. 1683-1698, 2020/11/07 2020.

[178] Z. Song *et al.*, "Study on the Effect of Cervical Spine Somatosensory Games of Virtual Reality and Augmented Reality on Relieving Neck Muscle Fatigue," Cham, 2020, pp. 359-375: Springer International Publishing.

[179] I. A. C. Giglioli, L. A. Carrasco-Ribelles, E. Parra, J. Marín-Morales, and M. Alcañiz Raya, "An Immersive Serious Game for the Behavioral Assessment of Psychological Needs," vol. 11, no. 4, p. 1971, 2021.

[180] J. D. Hart, T. Piumsomboon, L. Lawrence, G. A. Lee, R. T. Smith, and M. Billinghurst, "Emotion Sharing and Augmentation in Cooperative Virtual Reality Games," presented at the Proceedings of the 2018 Annual Symposium on Computer-Human Interaction in Play Companion Extended Abstracts, Melbourne, VIC, Australia, 2018. Available: https://doi.org/10.1145/3270316.3271543